\documentclass[12pt, a4paper]{article}

\font\small=cmr10

\usepackage{amsmath}
\usepackage{amsfonts,amssymb}
\newcommand{\h}{{\mathsf{h}}}
\newcommand{\mi}{{\mathrm{i}}}

\newcommand{\set}[1]{\left\{ #1\right\}}
\newcommand{\braket}[2]{\left\langle #1, #2\right\rangle}  
\newcommand{\ketbra}[2]{\left\vert #1\right\rangle\left\langle #2\right\vert}  
\newcommand{\tr}[1]{{\rm tr }\left(#1\right)}  
\newcommand{\Tr}[1]{{\rm Tr }\left(#1\right)}

\newcommand{\bo}[1]{\mathcal{B}(#1)}

\newcommand\unit{\hbox{\rm 1\kern-2.8truept l}}
\newcommand{\Ll}{{\mathcal{L}}}
\newcommand{\Tt}{{\mathcal{T}}}

\newcommand{\avanti}{\overrightarrow{\Omega}}
\newcommand{\indietro}{\overleftarrow{\Omega}}
\newcommand{\deava}{\overrightarrow{D}}
\newcommand{\deind}{\overleftarrow{D}}
\newcommand{\eava}{\overrightarrow{E}}
\newcommand{\eind}{\overleftarrow{E}}
\newcommand{\Lava}{\overrightarrow{L}}
\newcommand{\Lind}{\overleftarrow{L}}

\newcommand{\phiava}{{\overrightarrow{\Phi}_*}}
\newcommand{\phiind}{{\overleftarrow{\Phi}_*}}

\newcommand{\supp}[1]{{{\rm supp}(#1)}}

\newcommand{\eprod}[1]{{\mathbf{ep}(#1)}}

\newtheorem{definition}{Definition}
\newtheorem{theorem} {Theorem}
\newtheorem{proposition}{Proposition}

\newtheorem{remark}{Remark}
\newtheorem{lemma} {Lemma}
\newtheorem{corollary}{Corollary}

\begin{document}
\title{\bf Entropy Production for Quantum Markov Semigroups}
%
%
%
%
\date{}
\maketitle
\author{}
\centerline{{F. Fagnola}} 
\centerline{\small{Dipartimento di Matematica, Politecnico di Milano}} 
\centerline{\small{Piazza Leonardo da Vinci 32, I-20133 Milano (Italy)}}
\centerline{\small{franco.fagnola@polimi.it}}
\bigskip
\centerline{{R. Rebolledo}} 
\centerline{\small{Centro de An\'alisis Estoc\'astico y Aplicaciones }} 
\centerline{\small{ Facultad de Ingenier{\'\i}a 
y Facultad de Matem\'aticas }} 
\centerline{\small{Pontificia Universidad Cat\'olica de Chile}} 
\centerline{\small{Casilla 306, Santiago 22, Chile.}}
\centerline{\small{rrebolle@uc.cl}}
\begin{abstract}
An invariant state of a quantum Markov semigroup is an  
equilibrium state if it satisfies a quantum detailed balance 
condition. In this paper, we introduce a notion of entropy 
production for faithful normal invariant states of a quantum 
Markov semigroup on $\bo{\h}$ as a numerical index measuring 
``how much far'' they are from equilibrium.
The entropy production is defined as derivative of the 
relative entropy of the one-step forward and backward 
evolution in analogy with  the classical probabilistic concept.  
We prove an explicit trace formula expressing the 
entropy production in terms of the completely positive 
part of the generator of a norm continuous quantum Markov 
semigroup showing that it turns out to be zero if and only if 
a standard quantum detailed balance condition holds.
\end{abstract}
%


\section{Introduction}

This paper proposes a novel perspective on non equilibrium 
dissipative evolution of open quantum systems within the 
Markovian approach.  In this context, equilibrium states are 
invariant states characterised by a quantum detailed balance 
condition (see \cite{Agarwal,Alicki76,DeFr,KFGV,Maje,Talkner}), 
a natural property generalising classical detailed balance. 
However, a concept that distinguishes, among non 
equilibrium states, those that on one hand have a rich non 
trivial structure and, on the other hand, are sufficiently 
simple to allow a detailed study, is still missing.

Entropy production has been proposed, in several papers
\cite{HeinzPeter:2003p34,MR2046709,Qian2003,JaksicPill,MRM,Onsager:1931p1766}
as an index of deviation from detailed balance related with 
a rate of entropy variation. 
In \cite{Fagnola-Rebolledo-QP29} we proposed a definition 
of entropy production for faithful normal invariant states of 
quantum Markov semigroups analogous those  for
classical Markov semigroups applied to model particle 
interaction in classical mechanics. 
The entropy production was defined as the derivative of the 
relative entropy of the one-step forward and backward two-point 
states (Definition \ref{def:entr-prod} here) obtained from 
a maximally entangled state deformed by means of the 
given invariant state (see (\ref{eq:density_Omega0})).

In this paper, we prove an explicit trace formula for the entropy 
production in terms of the completely positive part of the 
generator of a norm continuous quantum Markov semigroups 
(Theorem \ref{th:ep-formula}). 
Our formula shows that non zero entropy production is closely 
related with violation of quantum detailed balance conditions 
and points out states with finite entropy production as a rich class of simple non equilibrium invariant states. Moreover, it provides 
an operator analogue (Theorem \ref{th:support-state2} (a))
of a necessary condition for finiteness of classical  
entropy production in terms of transition intensities, namely 
$\gamma_{jk}>0$ if and only  $\gamma_{kj}>0$. 

The plan of the paper is as follows. In Section \ref{Sect:QDB} 
quantum detailed balance conditions are reviewed and the 
key result on the structure of generators is recalled. The 
forward and backward states two-point states are introduced  
in Section \ref{Sect:forward-backward-state} starting from 
quantum detailed balance conditions and their densities
are computed. Entropy production is defined in Section 
\ref{Sect:ep} and the explicit formula is proved in Section
\ref{sect:entropy-prod-formula}. Three examples 
illustrating how entropy production indicates deviation from 
detailed balance are presented in Section \ref{Sect:examples}.

Finally we discuss some features of our results and possible 
directions for further investigation.

\section{Quantum detailed balance conditions}
\label{Sect:QDB}

Let $\mathcal{A}$ be a von Neumann algebra with a faithful 
normal state $\omega$ and identity $\unit$. A quantum 
Markov semigroup (QMS) on $\mathcal{A}$ is  a
weakly$^*$-continuous semigroup $\Tt=(\Tt_t)_{t\ge 0}$ 
of normal, unital, completely positive maps on $\mathcal{A}$.  
The predual semigroup on $\mathcal{A}_*$ will be denoted 
by $\Tt_*=(\Tt_{*t})_{t\ge 0}$.

The state $\omega$ is invariant if $\omega(\Tt_t(a))=\omega(a)$ 
for all $a\in\mathcal{A}$ and $t\ge 0$. A number of conditions 
called \emph{quantum detailed balance} (QDB) conditions have 
been proposed in the literature to distinguish, among invariant 
states, those enjoying reversibility properties. 

The first one, to the best of our knowledge, appeared in  
the work of Agarwal \cite{Agarwal} in 1973.  
Later extended and studied in detail by Majewski \cite{Maje}, it 
involves a reversing operation  $\Theta:\mathcal{A}\to\mathcal{A}$, 
namely a linear $*$-map (\,$\Theta(a^*)= \Theta(a)^*$ for 
all $a\in\mathcal{A}$), that is also an antihomomorphism 
(\,$\Theta(ab)=\Theta(b)\Theta(a)$\,) and satisfies $\Theta^2=I$, 
where $I$ denotes the identity map on $\mathcal{A}$.
A QMS satisfies the Agarwal-Majewski QDB condition if 
$\omega\left( a\Tt_t(b)\right) = 
\omega\left(  \Theta(b)\Tt_t(\Theta(a))\right)$,
for all $a,b\in\mathcal{A}$. If the state $\omega$  is 
invariant under the reversing operation, i.e. 
$\omega(\Theta(a))=\omega(a)$ for all $a\in\mathcal{A}$, 
as we shall assume throughout the paper,
this condition can be written in the equivalent form 
$\omega\left( a\Tt_t(b)\right) = 
\omega\left( (\Theta\circ\Tt_t\circ\Theta)(a) b\right)$ 
for all $a,b\in\mathcal{A}$. 
Therefore the Agarwal-Majewski QDB condition means that 
maps $\Tt_t$  admit dual maps coinciding with 
$\Theta\circ\Tt_t\circ\Theta$ for all $t\ge 0$; in 
particular dual maps must be positive since 
$\Theta$ is obviously positivity preserving. The map 
$\Theta$ often appears in the physical literature 
(see e.g. Talkner \cite{Talkner} and the references therein) 
as a parity map; a self-adjoint $a$ is an even (resp. odd)
observable if $\Theta(a)=a$ (resp.  $\Theta(a)=-a$).

When $\mathcal{A}=\bo{\h}$, the von Neumann algebra 
of all bounded operators on a complex separable Hilbert 
space $\h$,  as it is often the case for open quantum systems, 
the typical $\Theta$ is given by $\Theta(a)=\theta a^*\theta$ 
where $\theta$ is the conjugation with respect to a fixed 
orthonormal basis $(e_n)_{n\ge 0}$ of $\h$ acting as 
\begin{equation}\label{eq:conjugation}
\theta\bigg( \sum_{n\ge 0} u_n e_n \bigg)
= \sum_{n\ge 0}\bar{u}_n e_n.
\end{equation}
The operator $\theta$, however, can be any antiunitary 
($\braket{\theta v}{\theta u}=\braket{u}{v}$ for all 
$u,v\in\h$) such that $\theta^2=\unit$. 
Moreover, from $\omega(\theta a^*\theta) 
= \omega(a)$, letting $\rho$ denote the density of $\omega$ 
and denoting by $\tr{\cdot}$ the trace on $\h$, 
the linear operator $\theta \rho\theta$ being self-adjoint by 
$\braket{v}{\theta\rho\theta u} = 
\braket{\rho\theta u}{\theta v}=
\braket{\theta u}{\rho\theta v}= 
\braket{\theta \rho \theta v}{u}$, we have 
\[
\tr{\rho a} =\tr{\rho \theta a^* \theta}
=\sum_n\braket{e_n}{\rho \theta a^* \theta e_n}
=\sum_n\braket{\theta\rho\theta a^* (\theta e_n)}{(\theta e_n)}
= \tr{\theta \rho \theta a}
\] 
for all $a\in\mathcal{A}$, thus $\rho=\theta\rho\theta$, 
i.e. $\theta$ commutes with $\rho$. This assumption is 
reasonable because $\rho$ is often a function of energy 
which is an even observable, therefore it applies throughout 
the paper.

The best known QDB notion, however, is due to Alicki 
\cite{Alicki76}, \cite{Alicki-Lendi} and Kossakowski, Frigerio, Gorini, 
Verri \cite{KFGV}.  According to these authors, the QDB holds if 
there exists a dual QMS $\widetilde{\Tt}=
\left(\widetilde{\Tt}_t\right)_{t\ge 0}$ on $\mathcal{A}$ such that 
$\omega\left(a\Tt_t(b)\right)
=\omega\left(\widetilde{\Tt}_t(a)b\right)$ and the difference 
of generators $\Ll$ and $\widetilde{\Ll}$ is a derivation. 

Both the above QDB conditions depend in a crucial way from the 
bilinear form $(a,b)\to \omega(a b)$.  Indeed, when they hold true,   
all positive maps $\Tt_t$ admit \emph{positive} dual maps; 
as a consequence, all the maps $\Tt_t$ must commute 
with the modular group $(\sigma^\omega_t)_{t\in\mathbb{R}}$ 
associated with the pair $(\mathcal{A},\omega)$ 
(see \cite{KFGV} Prop. 2.1, \cite{MaSt} Prop. 5). 
This algebraic restriction is unnecessary if we consider the
bilinear form $(a,b)\to \omega\left(\sigma_{\mi/2}(a)b \right)$
introduced by Petz \cite{Petz} in the study of Accardi-Cecchini  conditional expectations. In this way, as noted by Goldstein 
and Lindsay (see \cite{GoLi}, \cite{Cipriani}), one can define 
dual QMS,  also when maps $\Tt_t$ do not commute with the 
modular group. Dual QMS defined in this way are called 
KMS-duals in contrast with GNS-duals defined via the bilinear 
form $(a,b)\to \omega\left(a b \right)$.

QDB conditions arising when we consider KMS-duals instead 
of GNS-duals are called \emph{standard} (see e.g. \cite{DeFr}, 
\cite{FFVU});  we could not find them in the literature, but  
it seems that they belong to the folklore of the subject. 
In particular, they were considered by R. Alicki and 
A. Majewski (private communication). 

\begin{definition}\label{def:SQDB}
Let $\Tt$ be a QMS with a dual QMS $\Tt^\prime$  satisfying 
$\omega\left(\sigma_{\mi/2}(a)\Tt_t(b) \right) =\omega
\left(\sigma_{\mi/2}\left(\Tt_t^\prime(a)\right)b \right)$
for all  $a,b\in\mathcal{A}$, $t\ge 0$. The semigroup $\Tt$ 
satisfies:
\begin{enumerate}
\item the standard quantum detailed balance 
condition with respect to the reversing operation $\Theta$ 
(SQBD-$\Theta$) if $\Tt_t^\prime= 
\Theta\circ\Tt_t\circ\Theta$ for all $t\ge 0$,
\item the standard quantum detailed balance condition (SQDB) 
if the difference of generators $\Ll -\Ll^\prime$ of $\Tt$ and 
$\Tt^\prime$ is a densely defined derivation.
\end{enumerate}
\end{definition}

It is worth noticing here that the above \emph{standard} QDB 
conditions coincide with the Agarwal-Majewski and Alicki-Gorini-Kossakowski-Frigerio-Verri respectively when the  
QMS $\Tt$ commutes with the modular group 
$(\sigma_t)_{t\in\mathbb{R}}$ associated with the pair 
$(\mathcal{A},\omega)$ (see, e.g.,  \cite{Cipriani,MaSt} 
and \cite{FFVU07,FFVU} for $\mathcal{A}=\bo{\h}$).

In the present paper we concentrate on QMS on $\bo{\h}$ which   
are the most frequent for open quantum systems. All states will 
be assumed to be normal and identified with their densities.
In particular, $\omega(x)=\tr{\rho\, x}$, $\sigma_t(x)= 
\rho^{\mi t} x \rho^{-\mi t}$ and the KMS duality reads
\begin{equation}\label{eq:KMS-duality}
\tr{\rho^{1/2}\, a\,  \rho^{1/2}\,  \Tt_t(b)}
=\tr{\rho^{1/2}\,  \Tt^\prime_t(a) \,  \rho^{1/2} b}.
\end{equation}
The map $\Theta$ will be the reversing operation 
$\Theta(x) = \theta x^* \theta$ where $\theta$ is the 
antiunitary conjugation (\ref{eq:conjugation}) with 
respect to some basis and the $\Tt$-invariant state $\rho$  
\emph{will be assumed to commute with $\theta$}. 
A Gram-Schmidt process shows that it is always possible 
to find such an orthonormal basis $(e_j)_{j\ge 1}$ of $\h$ 
of eigenvectors of $\rho$ that are also $\theta$-invariant 
(see Proposition \ref{prop:onb-theta-invariant} here).

First we recall the well-known  result (\cite{Partha} Theorem 30.16).

\begin{theorem}\label{th-special-GKSL} 
Let $\Ll$ be the generator of a norm-continuous QMS on 
$\bo{\h}$ and let $\rho$ be a normal state on $\bo{\h}$. There exists 
a bounded self-adjoint operator $H$ and a finite or infinite 
sequence $(L_\ell)_{\ell\ge 1}$ of elements of $\bo{\h}$ such that: 
\begin{enumerate}
\item[{\rm (i)}] $\hbox{\rm tr}(\rho L_\ell)=0$ for all $\ell\geq 1$, 
\item[{\rm (ii)}] $\sum_{\ell\geq 1}L^*_\ell L_\ell$ is a strongly 
convergent sum,
\item[{\rm (iii)}] if $(c_\ell)_{\ell\geq 0}$ is a square-summable 
sequence  of complex scalars and 
$c_0\unit+\sum_{\ell \geq 1}c_\ell L_\ell=0$ 
then $c_\ell=0$ for all $\ell\geq 0$,
\item[{\rm (iv)}] the following representation of $\Ll$ holds
\begin{equation}\label{eq:GKSL}
\Ll(x)=\mi[H,x] - \frac{1}{2}
\sum_{\ell\ge 1}
\left(L^*_\ell L_\ell x-2L^*_\ell xL_\ell + xL^*_\ell L_\ell\right)
\end{equation}
\end{enumerate}
If $H^\prime,(L^\prime_\ell)_{\ell\ge 1}$ is another family 
of bounded operators in $\bo{\h}$ with $H^\prime$ self-adjoint 
and the sequence $(L^\prime_\ell)_{\ell\ge 1}$ is finite or 
infinite, then the conditions {\rm (i)--(iv)} are fulfilled with 
$H,(L_\ell)_{\ell\ge 1}$  replaced by $H^\prime, 
(L^\prime_\ell)_{\ell\ge 1}$ respectively if and only if 
the lengths of the sequences $(L_\ell)_{\ell\ge 1}$, 
$(L^\prime_\ell)_{\ell\ge 1}$ are equal and for some scalar 
$c\in\mathbb{R}$ and a unitary matrix $(u_{\ell j})_{\ell j}$ 
we have 
\[
H^\prime=H+c,\qquad L^\prime_\ell=\sum_{j}u_{\ell j}L_j.
\]
\end{theorem}

Formula (\ref{eq:GKSL}) with operators $L_\ell$ satisfying (ii) 
and $H$ self-adjoint gives a GKSL 
(Gorini-Kossakowski-Sudarshan-Lindblad) representation of $\Ll$. 
A GKSL representation of $\Ll$ by means of operators $L_\ell,H$
satisfying also conditions (i) and (iii) will be called {\em special}.   

As an immediate consequence of uniqueness (up to a scalar) of 
the Hamiltonian $H$, the decomposition of $\Ll$ as the sum of the 
derivation $\mi[H,\cdot]$ and a dissipative part $\Ll_0
=\Ll-\mi[H,\cdot\,]$ 
determined by special GKSL representations of $\Ll$ is unique. 
Moreover, since $(u_{\ell j})$ is unitary, we have 
\[
\sum_{\ell\ge 1}\left(L^\prime_\ell\right)^*L^\prime_\ell 
=\sum_{\ell,k,j\ge 1}\bar{u}_{\ell k}u_{\ell j}L_k^*L_j 
=\sum_{k,j\ge 1}
\left(\sum_{\ell\ge 1}\bar{u}_{\ell k}u_{\ell j}\right) L_k^*L_j 
=\sum_{k\ge 1}L_k^*L_k. 
\]
Therefore, putting $G=-2^{-1}\sum_{\ell\ge 1}L^*_\ell L_\ell -iH$, 
we can write $\Ll$ in the form
\begin{equation}\label{eq:GKSL-G}
\Ll(x)= G^* x+
\sum_{\ell\ge 1} L^*_\ell xL_\ell + x G
\end{equation} 
where $G$ is uniquely determined by $\Ll$ up to a purely 
imaginary multiple of the identity operator. 

The unitary matrix $(u_{\ell j})_{\ell j}$ can obviously be realised 
as a unitary operator on a Hilbert space $\mathsf{k}$, called the  \emph{multiplicity space} with Hilbertian dimension equal to the 
length of the sequence $(L_\ell)_{\ell \ge 1}$ which is also 
uniquely determined by $\Ll$ by the minimality condition (iii).

\medskip
In \cite{FFVU} (Theorems 5, 8 and Remark 4) we proved the 
following characterisations of QMS satisfying a standard QDB 
condition.

\begin{theorem}\label{th:SQDB} 
A QMS $\Tt$ satisfies the SQDB if and only if for any special 
GKSL representation of the generator $\Ll$ by means of 
operators $G,L_\ell$ there exists a unitary $(u_{m\ell})_{m\ell}$ 
on $\mathsf{k}$ which is also symmetric (i.e. $u_{m\ell}=
u_{\ell m}$ for all $m,\ell$) such that, for all $k\ge 1$,
\begin{equation}\label{sqdb-cond}
\rho^{1/2}L^*_k=\sum_\ell u_{k\ell}L_\ell\rho^{1/2}.
\end{equation}
\end{theorem}

\begin{theorem}\label{th:SQDB-TR}
A QMS $\Tt$ satisfies the {SQBD-$\Theta$} condition if and only 
if for any special GKSL representation of $\Ll$ by means   
of operators $G, L_\ell$, there exists a self-adjoint unitary 
$(u_{k j})_{kj}$ on $\mathsf{k}$ such that:
\begin{enumerate}
\item \label{sdb-theta-1} $\rho^{1/2}\theta{G^*}\theta
                =G\rho^{1/2} $, 
\item \label{sdb-theta-2} $\rho^{1/2}\theta{L}_k^*\theta 
      = \sum_j u_{k j} {L_j}\rho^{1/2}$ for all $k\ge 1$.
\end{enumerate}
\end{theorem}
 
The SQBD-$\Theta$ condition is more restrictive than 
the SQDB condition because it involves also the identity 
$\rho^{1/2}\theta{G^*}\theta=G\rho^{1/2} $ (see 
Example \ref{ex:two-level}).
However,  this does not happen if $\theta G^* \theta = G$ 
and $\rho$ commutes with $G$. This is a reasonable physical 
assumption satisfied by many QMS as, for instance,  those
arising from the stochastic limit (e.g. \cite{AcLuVo,DeFr}).


The following result shows that, condition 2 alone, 
only implies that the difference 
$\Ll^\prime -  \Theta \circ \Ll \circ\Theta$ 
is a derivation (as in Alicki et al. QDB conditions) and 
clarifies differences between  Theorems \ref{th:SQDB} 
and \ref{th:SQDB-TR}.

\begin{theorem}\label{th:new-SQDB}
Let $\Tt$ be a QMS with generator $\Ll$ in a special GKSL form 
by means of operators $G$, $L_\ell$. Assume that 
$\rho^{1/2}\theta{L}_k^*\theta = \sum_j u_{k j} {L_j}\rho^{1/2}$,  
for all $k\ge 1$, for a self-adjoint unitary $(u_{k j})_{kj}$ on $\mathsf{k}$. Then 
\begin{equation}\label{eq:superQDB}
\Ll^\prime(x) - \left(\Theta \circ \Ll \circ\Theta\right)(x) 
= \mi \left[ K, x\right]
\end{equation}
with $K$ self-adjoint commuting with $\rho$.
\end{theorem}

\noindent{\it Proof.} Let $\Tt^\prime$ be the dual QMS of $\Tt$ 
as in (\ref{eq:KMS-duality}). Since
\[
\Ll^\prime(x) = 
\rho^{-1/2}\Ll_*\left(\rho^{1/2}x 
\rho^{1/2} \right)\rho^{-1/2},
\]
comparing special GKSL of $\Ll$ and $\Ll^\prime$ 
(as in \cite{FFVU}  Theorem 4), given a   special 
GKSL representation  of $\Ll$ we can find a special GKSL representation of $\Ll^\prime$ by means of $G^\prime$, $L_\ell^\prime$ such that 
\begin{equation}\label{eq:compare-GKSL-L-Lprime}
G^\prime = \rho^{1/2} G^* \rho^{-1/2}, 
\qquad
L_\ell^\prime = \rho^{1/2}  L_\ell^* \rho^{-1/2}.
\end{equation}
By condition (\ref{sdb-theta-2}.)  of Theorem 
\ref{th:SQDB-TR} and unitarity of $(u_{\ell k})_{\ell k}$ 
we have
\begin{eqnarray*}
\sum_{\ell}L_\ell^{\prime *} x L_\ell^\prime 
& = & \sum_{\ell }
\rho^{-1/2}  L_\ell \rho^{1/2}
x \rho^{1/2}  L_\ell^*  \rho^{-1/2}\\
& = &  \sum_{\ell, j,k} \bar{u}_{\ell j} u_{\ell k}
\theta L_j^* \theta x  \theta L_k \theta \\
& = & \sum_k \theta L_k^* \theta x  \theta L_k \theta.
\end{eqnarray*}
 It follows that 
$\Ll^\prime$ admits the special GKSL representation
\begin{equation}\label{eq:cp-parts-of-Ll-Llprime-coincide}
\Ll^\prime(x) = G^{\prime*} x + \sum_{\ell}
\theta L_\ell^* \theta x \theta L_\ell \theta + x G^\prime
\end{equation}
by means of $G^\prime$ and the operators $\theta L_k \theta $.

We now check that $G^\prime-\theta G\theta $ is 
anti-selfadjoint. Clearly, by the first identity  
(\ref{eq:compare-GKSL-L-Lprime}), it suffices to 
check that $\rho^{1/2}\left( G^\prime 
- \theta G \theta\right)\rho^{1/2}=\rho G^* 
-\rho^{1/2}\theta G \theta \rho^{1/2}$ 
is anti-selfadjoint.
The state $\rho$ is an invariant state for $\Tt_*$, thus 
$\Ll_*(\rho)=0$. The duality (\ref{eq:KMS-duality}) with 
$b=\unit$ shows that $\rho$ is also invariant for $\Tt^\prime_*$,  
then $ \Ll_{*}^\prime(\rho)=0$, and we find
from (\ref{eq:cp-parts-of-Ll-Llprime-coincide}) 
\[
\rho G^* + G\rho = \rho G^{\prime *} +G^\prime\rho 
= \rho^{1/2}\theta G\theta \rho^{1/2}
+ \rho^{1/2}\theta G^*\theta \rho^{1/2}
\]
namely 
\[
\rho G^* - \rho^{1/2}\theta G\theta\rho^{1/2}
= \rho^{1/2}\theta G^*\theta \rho^{1/2} - G\rho
=-\left(  \rho G^* - \rho^{1/2}\theta G\theta\rho^{1/2} \right)^*.
\]

It follows that $\Ll^\prime - \left(\Theta\circ\Ll \circ\Theta\right)
= \mi [K,\cdot]$ 
with $K$ selfadjoint commuting with $\rho$ since 
$\Ll_*(\rho)=\Ll^\prime_*(\rho)=0$.
{\hfill $\square$}

The SQDB condition without reversing operation (Definition  \ref{def:SQDB}. 2.) might be paralleled with reversing operation, requiring (\ref{eq:superQDB}), however, we could not find 
this QDB condition in the literature.

\section{Forward and backward two-point states} 
\label{Sect:forward-backward-state}

We now introduce the \emph{two-point forward and 
backward states}.


\begin{definition}\label{def:two-point-states}
The forward two-point state is the normal state on $\bo{\h}\otimes\bo{\h}$ given by
\begin{equation}\label{forward-state}
\avanti_t\left(a\otimes b\right)
=\tr{\rho^{1/2} \theta a^*\theta\rho^{1/2} \Tt_t(b)}, 
\qquad  a,b\in\bo{\h}; 
\end{equation}
the \emph{backward two-point state} is the normal 
state on on $\bo{\h}\otimes\bo{\h}$ given by
\begin{equation}\label{backward-state}
\indietro_t\left(a\otimes b\right)
=\tr{\rho^{1/2}\theta{\Tt_t(a^*)}\theta\rho^{1/2}b}, 
\qquad  a,b\in\bo{\h}.
\end{equation}
\end{definition}

It is clear that both $\avanti_t$ and $\indietro_t$ 
are normalised linear functionals on  $\bo{\h}\otimes\bo{\h}$ 
since $\theta(za)^*\theta = \theta \bar{z}a^*\theta
=z \theta a^*\theta$, for all $z\in\mathbb{C}$ and 
all $a\in \bo{\h}$. They are positive and normal by the 
following proposition also giving their densities.

\begin{proposition}\label{prop:density_Omega0} 
Let $\rho=\sum_j\rho_j\ketbra{e_j}{e_j}$ be a spectral
decomposition of $\rho$.
The density of states $\avanti_0=\indietro_0$ is  
the rank one projection  
\begin{equation}\label{eq:density_Omega0}
D=\ketbra{r}{r}, \qquad 
r= \sum_{j}\rho_j^{1/2} \theta e_j\otimes e_j
\end{equation}
The densities of the forward and backward states are 
respectively
\begin{equation}\label{eq:FBdensities}
\deava_t = (I\otimes \Tt_{*t})(D), \qquad 
\deind_t = (\Tt_{*t}\otimes I)(D).
\end{equation} 
\end{proposition}

\noindent{\it Proof.} For all $a,b\in\bo{\h}$ we have
\begin{eqnarray*}
\left\langle r, (a\otimes b)r \right\rangle
 & = & \sum_{j,k}\left(\rho_j\rho_k\right)^{1/2}
\left\langle \theta e_j\otimes e_j, 
  (a\otimes b)\theta e_k\otimes e_k\right\rangle\\
& = & \sum_{j,k}\left(\rho_j\rho_k\right)^{1/2}
\left\langle \theta e_j, a \theta e_k\right\rangle
\left\langle e_j, b e_k\right\rangle\\
& = & \sum_{j,k}\left(\rho_j\rho_k\right)^{1/2}
\left\langle \theta a\theta e_k, e_j\right\rangle
\left\langle e_j, b e_k\right\rangle\\
& = & \sum_{k}  \rho_k ^{1/2}
\left\langle \theta a\theta e_k, \rho^{1/2}b e_k\right\rangle\\
& = & \sum_{k}  
\left\langle \theta a\theta \rho^{1/2}e_k, \rho^{1/2}b e_k\right\rangle\\
& = & \tr{  \rho^{1/2}\theta a^*\theta\rho^{1/2}b }.
\end{eqnarray*}
Formulae (\ref{eq:FBdensities}) follow immediately 
from 
\[
\avanti_t\left(a \otimes b\right) 
= \avanti_0\left(a \otimes \Tt_t(b)\right), \qquad
\indietro_t\left(a \otimes b\right) 
= \indietro_0\left(\Tt_t(a) \otimes b\right).
\]
{\hfill $\square$}

The entropy production will be defined in the next section by 
means of the relative entropy of the forward and backward 
two-point states.

\begin{remark}\label{Choi-Jamiolkowski}{\rm 
Note that, when $\mathsf{h}=\mathbb{C}^d$ and 
$\theta e_j= e_j$ for all $j$, we have
\[
\ketbra{r}{r} = \left(\rho^{1/2}\otimes\unit\right)
\left(d^{-1}\sum_{j=1}^d
\ketbra{e_j\otimes e_j}{e_j\otimes e_j}\right)
\left(\rho^{1/2}\otimes\unit\right)
\]
(and the same formula replacing  $\rho^{1/2}\otimes\unit$ 
by $\unit\otimes\rho^{1/2}$).
Therefore $\ketbra{r}{r}$ may be viewed as a $\rho$ 
deformation of a maximally entangled state and $\deava_t$, 
$\deind_t$ are the image of $I\otimes \Tt_{*t}$, 
$\Tt_{*t}\otimes I$ under the Choi-Jamio{\l}kowski 
isomorphism.}
\end{remark}

\begin{remark}\label{rem:opposite-algebra}
{\rm
Operators $\theta x^*\theta$ can be thought of as elements of 
the \emph{opposite} algebra $\bo{\h}^{\rm o}$ of $\bo{\h}$. 
Indeed, recall that $\bo{\h}^{\rm o}$ is in one-to-one 
correspondence with $\bo{\h}$ as a set via the trivial 
identification $x\to x^{\rm o}$, has the same vector space 
structure, involution and norm but the product $\circledcirc$ 
is given by $ x^{\rm o}\circledcirc y^{\rm o} = (yx)^{\rm o}$.
Therefore, the linear map $\Theta:\bo{\h}\to\bo{\h}^{\rm o}$ 
defined by $x\to \theta x^*\theta$ is a $^*$-isomorphism of 
$\bo{\h}$ onto $\bo{\h}^{\rm o}$ since
\[
\Theta(x)\circledcirc\Theta(y)
=\theta x^*\theta \circledcirc \theta y^*\theta 
= \theta y^*\theta \theta x^*\theta 
= \theta (xy)^*\theta
=\Theta(xy) .
\]
Clearly $\Theta\otimes I:\bo{\h}\otimes \bo{\h}\to 
\bo{\h}^{\rm o}\otimes \bo{\h}$ is a $^*$-isomorphism.
This remark is useful for defining entropy production as 
an index measuring deviation from standard detailed balance 
\emph{without} time reversal in a similar way. One can define 
the state $\avanti_0^\prime=\indietro_0^\prime$ on
$\bo{\h}^{\rm o}\otimes \bo{\h}$ by 
\[
\avanti_0^\prime(x\otimes y)
=\tr{\rho^{1/2}x\rho^{1/2}y}
\]
Note that element $Z$ of  $\bo{\h}^{\rm o}\otimes \bo{\h}$ 
is ``positive'' if and only if $(\Theta\otimes I)(Z)$ is positive in 
$\bo{\h}\otimes \bo{\h}$ because $\Theta\otimes I$ is a 
$^*$-isomorphism and $(\Theta\otimes I)^2$ is the identity 
map.

We can define the entropy production again considering the 
relative entropy of $\deava_t$ and $\deind_t$ but now viewed 
as densities of states on $\bo{\h}^{\rm o}\otimes \bo{\h}$. 
}
\end {remark}

We finish this section with a couple of useful properties of $r$.

\begin{proposition}\label{prop:cyclic-separating}
The vector $r$ is cyclic and separating for subalgebras 
$\unit\otimes \bo{\h}$ and $\bo{\h}\otimes \unit$. 
\end{proposition}

\noindent{\it Proof.} Let $X\in\bo{\h}$ and let $\rho=
\sum_j\rho_j\ketbra{ e_j}{e_j}$ 
be a spectral decomposition of $\rho$. Then  
$(\unit\otimes X)r =0$ if and only if  
$\sum_j \rho_{j}\, \theta e_j\otimes (Xe_j)=0$, i.e. 
$Xe_j=0$ for all $j$ since $\rho_j>0$ and vectors 
$\theta e_j$ are linearly independent.
It follows that $X=0$. 

The same argument shows that $r$ is also separating for 
$\bo{\h}\otimes \unit$. Therefore it is cyclic for 
$\unit\otimes \bo{\h}$ and $\bo{\h}\otimes \unit$ because 
these subalgebras of $\bo{\h}\otimes \bo{\h}$ are 
mutual commutants.
{\hfill $\square$}

\begin{proposition}\label{prop:r-ortogonale-Xr}
An operator $X\in \bo{\h}$ satisfies $\tr{\rho X}=0$ if and 
only if  $(\unit\otimes X)r$ and $(X\otimes \unit)r$ are 
orthogonal to $r$ in $\h\otimes\h$.
\end{proposition}

\noindent{\it Proof.} Immediate from
$ \braket{r}{(\unit\otimes X)r} =\braket{r}{(X\otimes \unit)r}
=\tr{\rho X}$.
{\hfill $\square$}

\section{Entropy production for a QMS}
\label{Sect:ep}

In the sequel $\Tr{\cdot}$ denotes 
the trace on $\h\otimes\h$.

The \emph{relative entropy} of $\avanti_t$ with respect 
to $\indietro_t$ is given by
\[
S\left(\avanti_t,\indietro_t\right)
=\Tr{\deava_t\left(\log \deava_t-\log \deind_t\right)},
\]
if the support of $\avanti_t$ is included in that of $\indietro_t$ 
and  $+\infty$ otherwise. 

\begin{definition}\label{def:entr-prod}
The \emph{entropy production rate} of a QMS  $\Tt$ and 
invariant state $\rho$ is defined by 
\begin{equation}
\eprod{\Tt,\rho}=\limsup_{t\to 0^+}
\frac{S\left(\avanti_t,\indietro_t\right)}{t}
\end{equation}
\end{definition}

\begin{remark} {\rm
The entropy production (entropy production for short) 
$\eprod{\Tt,\rho}$ is clearly non-negative. It coincides 
with the right derivative of 
$S\left(\avanti_t,\indietro_t\right)$ at $t=0$, if the 
limit exists, since $S\left(\avanti_0,\indietro_0\right)=0$.  
Moreover, $\eprod{\Tt,\rho}$ vanishes if the SQBD-$\Theta$ 
(or the SQDB viewing $\avanti_t$ and $\indietro_t$ as states 
on $\bo{\h}^{\rm o}\otimes \bo{\h}$) holds.

Under the assumptions of Theorem \ref{th:ep-formula}, the 
entropy production formula (\ref{eq:ep-formula}) we are going 
to prove, shows that, if $\eprod{\Tt,\rho}=0$, then the SQDB 
condition holds as well as the  SQBD-$\Theta$  condition under 
if  $\theta G^*\theta=G$ and $\rho\theta=\theta\rho$. 
A counterexample in subsection \ref{ex:two-level} 
shows that  SQBD-$\Theta$ may fail without these 
commutation assumptions even if $\eprod{\Tt,\rho}$ is zero.

Our definition gives a true non-commutative analogue of 
entropy production for classical Markov semigroups \cite{Qian2003}. 
We refer to \cite{Fagnola-Rebolledo-QP29} subsection 2.2 for 
a detailed discussion.
}
\end{remark}

\begin{proposition}\label{prop:state-derivative}
Let $\deava_t$ and $\deind_t$ be the densities of the forward 
and backward two-point states as in (\ref{eq:FBdensities}). 
The following are equivalent:
\begin{itemize}
\item [(a)] $\deava_t=\deind_t$, for all $t\geq 0$,
\item [(b)] $ (I \otimes\Ll_{*})(D)=(\Ll_{*}\otimes I)(D)$.
\end{itemize}
\end{proposition}

\noindent{\it Proof.}
Clearly (a) implies (b) by differentiation at time $t=0$.

Conversely, if (b) holds, since $I\otimes{\Ll}_{*}$ and ${\Ll}_{*}\otimes I$ commute, we have
\[
(I\otimes{\Ll}_{*})^2(D) = 
(I\otimes{\Ll}_{*})({\Ll}_{*}\otimes I )(D)
=({\Ll}_{*}\otimes I)( I \otimes{\Ll}_{*})(D)
=({\Ll}_{*}\otimes I)^2(D).
\]
Thus, by induction, we find $(I  \otimes{\Ll}_{*})^n(D)
=({\Ll}_{*}\otimes I)^n(D)$, for all $n\ge 1$, so that
\[
\deava_t=\sum_{n\geq 0}\frac{t^n}{n!}
(I\otimes{\Ll}_{*})^n(D)
=\sum_{n\geq 0}\frac{t^n}{n!}
({\Ll}_{*}\otimes I)^n(D)
=\deind_t,
\]
for all $t\geq 0$ and (a) is proved.
{\hfill $\square$}

The following proposition shows, in particular, that the relative 
entropy of the forward and backward two-point state is 
symmetric.

\begin{proposition}\label{prop:EP-symm}
The  relative entropy of $\avanti_t$ with respect 
to $\indietro_t$ satisfies
\begin{equation}\label{eq:EP-symm}
S\left(\avanti_t,\indietro_t\right)=\frac{1}{2} \Tr{\left(\deava_t-\deind_t\right)
\left(\log\deava_t-\log\deind_t\right)}.
\end{equation}
In particular, if $S\left(\avanti_t,\indietro_t\right)$ is finite, 
then the densities $\deava_t$, $\deind_t$ have the same 
support.
\end{proposition}

\noindent{\it Proof.} Let $F$ be the unitary flip operator on $\h\otimes\h$ defined by $F e_j\otimes e_k= e_k\otimes e_j$. 
Noting that $F \deava_t F = \deind_t$ and then 
$F \log\left(\deava_t\right) F = \log\left(\deind_t\right)$, 
we have 
\begin{eqnarray*}
S\left(\avanti_t,\indietro_t\right)
& = &\Tr{F\deava_t\left(\log \deava_t-\log \deind_t\right)F} \\
& = & \Tr{-\deind_t\left(\log \deava_t-\log \deind_t\right)}
\end{eqnarray*}
Therefore
\[
2S\left(\avanti_t,\indietro_t\right)
= \Tr{\deava_t\left(\log \deava_t-\log \deind_t\right)}
+\Tr{-\deind_t\left(\log \deava_t-\log \deind_t\right)}
\]
and (\ref{eq:EP-symm}) follows. 

If $S\left(\avanti_t,\indietro_t\right)$ is finite, then the support 
$\supp{\deava_t}$ of $\deava_t$ is contained in the support  $\supp{\deind_t}$ of $\deind_t$. By the identity  $F \deava_t F = \deind_t$, we have then
\[
\supp{\deind_t}=F\,\supp{\deava_t}F 
\subseteq F\,\supp{\deind_t}F = \supp{\deava_t},
\] 
and the proof is complete.
{\hfill $\square$}

Proposition \ref{prop:EP-symm} shows that the first step 
towards the computation of the entropy production is  to
check if $\deava_t$ and $\deind_t$ have the same support
for $t$ in a right neighbourhood of $0$. This is a somewhat 
technical point (as in the classical case \cite{Qian2003}) 
if both $\deava_t$ and $\deind_t$ do not have full support. 
In Section \ref{sect:support-problems} we develop a  
simple method for solving this problem.

\section{Entropy production formula}
\label{sect:entropy-prod-formula}

In this section we establish our entropy production formula 
under the following assumption on supports of the forward 
and backward state.

\smallskip
\noindent({\bf FBS}) \ {\it Supports of $\deava_t$ and 
$\deind_t$ coincide and are finite dimensional.}
\smallskip

Finite dimensionality is needed for the application of results in perturbation theory. 
Supports of $\deava_t$ and $\deind_t$ may vary 
with $t$ even if they coincide and are finite dimensional. 
A simple example arises when we consider a semigroup 
$(\Tt_t)_{t\ge 0}$ of automorphisms of $\bo{\h}$ with 
$L_\ell=0$ for all $\ell$ and a  non-zero self-adjoint 
operator $H$. Any faithful density $\rho$ commuting with $H$ 
provides a faithful invariant state.

Let $\phiava$ and $\phiind$ be the linear maps on 
trace class operators on $\h\otimes\h$
\begin{equation}\label{eq:phiava-phiind}
\phiava(X)= \sum_{\ell}\left(\unit\otimes L_\ell\right) 
X \left(\unit\otimes L_\ell^*\right), \qquad
\phiind(X) = \sum_{\ell}\left(L_\ell \otimes \unit\right) 
X \left( L_\ell^*\otimes \unit\right)
\end{equation}
where $L_\ell$ are the operators of a special GKSL representation 
of $\Ll$. Recall that, by Proposition \ref{prop:r-ortogonale-Xr}, 
$\left(\unit\otimes L_\ell\right) r$ and 
$\left(L_\ell \otimes \unit\right)$ are orthogonal to $r$.

\begin{theorem}\label{th:ep-formula} 
Let $\Tt$ be a norm continuous QMS on $\bo{\h}$ with a 
faithful, normal invariant state $\rho$. Under the assumption {\rm({\bf FBS})} the entropy production is
\begin{equation}\label{eq:ep-formula}
\eprod{\Tt,\rho} = \frac{1}{2}
\Tr{\left(\phiava(D)-\phiind(D)\right)
\left(\log\left(\phiava(D)\right) 
- \log\left(\phiind(D)\right)
\right)}.
\end{equation}
\end{theorem}

The rest of this section is devoted to proving 
(\ref{eq:ep-formula}). 

Let $S_t$ denote this common finite dimensional ($k+1$ 
dimensional, say) support of $\deava_t$ and $\deind_t$. 
Since $\deind_t=F\deava_t F$, for all $t$, we 
can write spectral decompositions
\begin{equation}\label{eq:spectral-dec-deava-deind}
\deava_t = \sum_{\ell =0}^k \lambda_\ell(t) \eava_\ell(t), 
\qquad 
\deind_t = \sum_{\ell =0}^k \lambda_\ell(t) \eind_\ell(t), 
\end{equation}
where $\lambda_\ell(t)$ are common eigenvalues and 
all spectral projections satisfy
\[
\eind_\ell(t) = F\, \eava_\ell(t)\, F
\]
for all $t\ge 0$. Moreover, since $S_t$ is $(k+1)$-dimensional 
for all $t>0$, we have $\lambda_\ell(t)>0$ for all $t>0$ and 
$\ell=0,1,\dots, k$.

It is well known that, by deep results in finite-dimensional 
perturbation theory, Rellich's theorem and its consequences 
(see e.g.  Kato\cite{Kato66}, Theorem 6.1 p. 120, 
Reed and Simon\cite{RSv4} Theorems XII.3 p. 4,  
XII.4 p. 8 and concluding remark),  that we can choose 
\[
t \to \lambda_\ell(t), \qquad t\to \eava_\ell(t)
\]
as single-valued analytic functions of $t$ for $t$ in a 
neighbourhood of $0$. Moreover, noting that both 
$\deava_t$ and $\deind_t$ converge in trace norm 
to $D$  as $t$ tends to $0$ and $1$ is a simple eigenvalue 
of $D$, we can suppose, relabeling indexes if 
necessary, that 
\begin{equation}\label{eq:D-eigenvalue1}
\lim_{t\to 0}\lambda_0(t)=1, \qquad 
\lim_{t\to 0} \eava_0(t)=\lim_{t\to 0}\eind_0(t)= D.
\end{equation}

The difference $ \log\left(\deava_t\right)-\log\left(\deind_t\right)$ 
is a bounded operator on $S_t$  and we can define 
it as $0$ on the orthogonal complement of $S_t$. 
Moreover, denoting $\log\left(\deava_t\right)\big|_{S_t}$ and  
$\log\left(\deind_t\right)\big|_{S_t}$ restrictions to $S_t$, 
we can prove the following 

\begin{lemma}\label{lem:norm-le-logt}
There exists  constants $c>0, t_+>0$ and  
$m\in\mathbb{N}$ such that 
\[
\left\Vert \log\left(\deava_t\right)\big|_{S_t}\right\Vert 
\le c - m \log(t) , \qquad 
\left\Vert \log\left(\deind_t\right)\big|_{S_t}\right\Vert 
\le c -m\log(t) 
\]
for all $t\in\, ]0,t_+]$. 
\end{lemma}

\noindent{\it Proof.} Recall that functions $t\to\lambda_\ell(t)$ 
are analytic and strict positive in a right neighbourhood 
of $0$.  For each $\ell$, let $m_\ell$ be the order of the first 
non-zero (hence strictly positive) derivative of 
$t\to\lambda_\ell(t)$ at $t=0$. 
There exists $\varepsilon_\ell\in ]0,1[$ 
and $t_\ell>0$ such that $\lambda_\ell(t)\ge \varepsilon_\ell 
t^{m_\ell}$ for all $t\in ]0,t_\ell]$. Putting 
\[
\varepsilon= \min_{0\le \ell\le k}\varepsilon_\ell, \qquad 
m=\max_{0\le \ell\le k} m_\ell, \qquad 
t_+=\min_{0\le \ell\le k} t_\ell
\]  
we find then the inequality  
$ \lambda_\ell(t) \ge \varepsilon_\ell t^{m_\ell} \ge 
\varepsilon\, t^m $
for all $\ell$ and $t\in ]0,t_+]$. Therefore we have 
\[
\deava_t\big|_{S_t}\ge \varepsilon t^m \unit_{S_t}
\] 
where $\unit_{S_t}$ is the orthogonal projection onto $S_t$, 
and the norm estimate follows.

The proof for $\deind_t$ is identical.
{\hfill $\square$}

We now start computing the limit of 
\begin{equation}\label{eq:ep-limit01}
t^{-1}\Tr{\left(\deava_t-\deind_t\right)
\left(\log\deava_t-\log\deind_t\right)}
\end{equation}
for $t\to 0^+$.  As a first step note that 
\[
\lim_{t\to 0^+}t^{-1} \left(\deava_t-\deind_t\right) 
=  (I\otimes \Ll_*)(D)-(\Ll_*\otimes I)(D) 
\]
in trace norm. Moreover, denoting $\left\Vert\cdot\right\Vert_1$ 
the trace norm 
\[
\left\Vert t^{-1} \left(\deava_t-\deind_t\right) - 
\left((I\otimes \Ll_*)(D)-(\Ll_*\otimes I)(D)\right) \right\Vert_1
\]
is infinitesimal of order at most $t$ for $t$ tending to $0$, 
therefore the modulus of the difference of (\ref{eq:ep-limit01}) 
and 
\begin{equation}\label{eq:ep-limit02}
\Tr{\left((I\otimes \Ll_*)(D)-(\Ll_*\otimes I)(D)\right)
\left(\log\deava_t-\log\deind_t\right)}, 
\end{equation}
by Lemma \ref{lem:norm-le-logt} is not bigger than a constant 
times $(c-m\log(t))t$ and goes to $0$ for $t$ tending to $0^+$.

It suffices then to compute the limit of (\ref{eq:ep-limit02}) for 
$t$ tending to $0^+$.

We first analyse the behaviour of the $0$-th term of  
(\ref{eq:spectral-dec-deava-deind}). 

\begin{lemma}\label{eq:deava-deind-0th} 
The following limits hold:
\begin{eqnarray}
\lim_{t\to 0^+} t^{-1}\left( \lambda_0(t)\eava_0(t)-D\right) 
& = & \ketbra{(\unit\otimes G)r}{r}+ \ketbra{r}{(\unit\otimes G)r}\\
\lim_{t\to 0^+} t^{-1}\left( \lambda_0(t)\eind_0(t)-D\right) 
& = & \ketbra{(G\otimes\unit)r}{r}+ \ketbra{r}{(G\otimes\unit)r}
\end{eqnarray}
\end{lemma}

\noindent{\it Proof.} The proof is the same for $\eava_0(t)$ 
and $\eind_0(t)$, therefore we consider $\eava_0(t)$ dropping
the arrows and writing $\Ll_*(D)$ instead of 
$(I\otimes \Ll_*)(D)$  for notational convenience.

Let $t_0>0$ be sufficiently small such that $D_t$ has only the simple
eigenvalue $\lambda_0(t)$ in $[3/4,1]$ and all other eigenvalues 
in $[0,1/4]$ for all $t\in [0,t_0[$.
By well known formulae (see e.g. \cite{Kato66} Ch. I) for 
spectral projections, for $t$ small enough we have
\begin{eqnarray*}
E_0(t) = \frac{1}{2\pi \mi}\int_{C}\left(\zeta-D_t\right)^{-1}d\zeta,  
& \qquad & 
D = \frac{1}{2\pi \mi}\int_{C}
   \zeta\left(\zeta-D\right)^{-1}d\zeta,\\
\lambda_0(t) E_0(t) & = & \frac{1}{2\pi \mi}\int_{C}
   \zeta\left(\zeta-D_t\right)^{-1}d\zeta
\end{eqnarray*}
where $C$ is the circle $\{ z\in \mathbb{C} \, \mid \, |z-1|=1/2\,\}$. 
Therefore we can write
\begin{equation}\label{eq:lambda0E0menD-integrand}
\frac{\lambda_0(t)E_0(t)-D}{t}
=\frac{1}{2\pi \mi}\int_{C}
   \frac{\left(\zeta-D_t\right)^{-1}-\left(\zeta-D\right)^{-1}}{t}
\,\zeta\, d\zeta
\end{equation}
Note that, for all $t\in  ]0,t_0[$ 
\[
t^{-1}\left(\left(\zeta-D_t\right)^{-1}-\left(\zeta-D\right)^{-1}\right)
=t^{-1}\left(\zeta-D_t\right)^{-1}
\left( D_t - D\right)\left(\zeta-D\right)^{-1}
\]
implying the norm estimate
\[
 t^{-1}\left\Vert \left(\zeta-D_t\right)^{-1}
-\left(\zeta - D\right)^{-1}\right\Vert _1 
\le \left\Vert t^{-1}\left(D_t-D\right)\right\Vert_1
\cdot \left\Vert \left(\zeta-D_t\right)^{-1}\right\Vert
\cdot\left\Vert \left(\zeta-D\right)^{-1}\right\Vert.
\]
Now, since the operators $\left(\zeta-D_t\right)^{-1}$ 
and $\left(\zeta-D\right)^{-1}$ are normal with discrete 
spectrum, contained in the union of the intervals $[0,1/4]$
and $[3/4,1]$ of the real axis, their norm 
is smaller than
\[
\sup_{\zeta\in C,\, x\in[0,1/4]\cup[3/4,1]}
\left| \zeta - x\right|^{-1} \le 4.
\] 
Moreover 
\[
\left\Vert \frac{D_t-D}{t}\right\Vert_1
=\frac{1}{t}\left\Vert \int_0^{t} 
  \Tt_{* s}(\Ll_*(D))ds \right\Vert_1 
\le \frac{1}{t} \int_0^{t} \left\Vert \Ll_*(D)\right\Vert_1 ds  
 = \left\Vert \Ll_*(D)\right\Vert_1,
\]
thus we have  
\[
t^{-1}\left\Vert \left(\zeta-D_t\right)^{-1}
-\left(\zeta-D\right)^{-1} \right\Vert \le 16 \left\Vert \Ll_*(D)\right\Vert_1.
\]
The integrand of (\ref{eq:lambda0E0menD-integrand}) 
converges to 
$\zeta\left(\zeta-D \right)^{-1}\Ll_*(D)
\left(\zeta-D \right)^{-1}$ for $t$ going to $0$
thus, by the dominated convergence theorem, we find
\begin{equation}\label{eq:lambda0E0menD-integrand2}
\lim_{t\to 0^+}\frac{\lambda_0(t)E_0(t) - D}{t}
= \frac{1}{2\pi \mi}\int_{C}
   \zeta\left(\zeta-D \right)^{-1}\Ll_*(D)\left(\zeta-D \right)^{-1} d\zeta.
\end{equation}

The proof of Lemma \ref{eq:deava-deind-0th} ends computing the right-hand side. First note that 
\[
 \frac{1}{2\pi \mi}\int_{C}\zeta
  \left\langle r, \left(\zeta-D \right)^{-1}\Ll_*(D)
 \left(\zeta-D \right)^{-1}r \right\rangle d\zeta
=  \frac{1}{2\pi \mi}\int_{C}
   \left\langle r, \Ll_*(D) r \right\rangle 
\frac{\zeta\,d\zeta}{(\zeta-1)^2} 
\]
with $\left\langle r, \Ll_*(D) r \right\rangle=2\Re \langle r, G r\rangle$ 
and  
\[
\frac{1}{2\pi \mi}\int_{C} 
\frac{\zeta\,d\zeta}{(\zeta-1)^2} 
= \frac{1}{2\pi \mi}\int_{C} 
\frac{(\zeta-1)\,d\zeta}{(\zeta-1)^2} 
+ \frac{1}{2\pi \mi}\int_{C} 
\frac{d\zeta}{(\zeta-1)^2}   
= \frac{1}{2\pi \mi}\int_{C} 
\frac{d\zeta}{\zeta-1} = 1
\]
so that 
\begin{equation}\label{eq:lambda0E0menD-integrand3}
\lim_{t\to 0^+}\frac{1}{2\pi \mi}\int_{C}
  \left\langle r, \left(\zeta-D \right)^{-1}\Ll_*(D)
 \left(\zeta-D \right)^{-1}r \right\rangle d\zeta
= 2\Re \langle r, G r\rangle.
\end{equation}
Second, for all vector $v$ orthogonal to $r$ we have
\begin{eqnarray*}
\frac{1}{2\pi \mi}\int_{C}\zeta
  \left\langle r, \left(\zeta-D \right)^{-1}\Ll_*(D)
 \left(\zeta-D \right)^{-1} v \right\rangle d\zeta
& = &\frac{1}{2\pi \mi}\int_{C} 
  \left\langle r, \Ll_*(D)v \right\rangle 
  \frac{d\zeta}{\zeta-1} \\
& = &  \left\langle r, \Ll_*(D)v \right\rangle 
= \left\langle Gr, v\right\rangle
\end{eqnarray*}
since $r$ is orthogonal to all $(\unit\otimes L_\ell)r$ 
and $(L_\ell\otimes\unit )r$, and, in a similar way, 
\[
\frac{1}{2\pi \mi}\int_{C}
  \left\langle v, \left(\zeta-D \right)^{-1}\Ll_*(D)
 \left(\zeta-D \right)^{-1}r \right\rangle d\zeta
=\left\langle v, Gr\right\rangle.
\]
Third, for all $v,u$ orthogonal to $r$ 
\[
\frac{1}{2\pi \mi}\int_{C}
  \left\langle v, \left(\zeta-D \right)^{-1}\Ll_*(D)
 \left(\zeta-D \right)^{-1}u \right\rangle d\zeta
=\frac{1}{2\pi \mi}\int_{C}
  \left\langle v, \Ll_*(D)u \right\rangle \frac{d\zeta}{\zeta}=0
\]
because $\zeta\to \zeta^{-1}$ is holomorphic on 
the half plane containing $C$.

This completes the proof.
{\hfill $\square$}

\begin{lemma}\label{lem:deava-deind-1st}
The following limits hold:
\[
\lim_{t\to 0^+}\sum_{\ell=1}^k t^{-1}\lambda_\ell(t) \eava_\ell(t)
= \phiava(D),  \qquad  
\lim_{t\to 0^+}\sum_{\ell=1}^k t^{-1}\lambda_\ell(t) \eind_\ell(t)
= \phiind(D)
\]
Moreover there exists a special GKSL representation of $\Ll$ 
such that $\lambda^\prime_\ell(0)=
\left\Vert\Lava_\ell r\right\Vert^2=
\left\Vert\Lind_\ell r\right\Vert^2$ for 
$\ell =1,\dots,d$ and
\[
\lim_{t\to 0^+} \eava_\ell(t) 
= \frac{\ketbra{\Lava_\ell\, r}{\Lava_\ell\, r}}
{\left\Vert \Lava_\ell\, r\right\Vert^2}, \qquad
\lim_{t\to 0^+} \eind_\ell(t) 
= \frac{\ketbra{\Lind_\ell\, r}{\Lind_\ell\, r}}
{\left\Vert \Lind_\ell\, r\right\Vert^2}
\]
for all $\ell=1,\dots, d$.
\end{lemma}

\noindent{\it Proof.} 
The first identities follow immediately from Lemma 
\ref{eq:deava-deind-0th} writing
\[
\sum_{\ell=1}^k t^{-1}\lambda_\ell(t) \eava_\ell(t)
= t^{-1}\left(\deava_t-D\right) -t^{-1}\left(\eava_0(t)-D\right)
\]
and recalling that $t^{-1}\left(\deava_t-D\right)$ converges to $(I\otimes\Ll_*)(D)$. Moreover, note the $d\times d$ matrix 
$C$ with $c_{jk}= \braket{\Lava_j r}{\Lava_k r}
=\braket{\Lind_j r}{\Lind_k r}$ is self-adjoint. Let 
$U=(u_{jk})_{1\le j,k\le d}$ be a $d\times d$ unitary matrix 
such that $U^*CU$ is diagonal and consider the new special 
GKSL representation of $\Ll$ obtained replacing the operators 
$L_\ell$ by $\sum_h u_{h \ell } L_h$. 
Now we have 
\[
\braket{\Lava_j r}{\Lava_k r}
=\braket{\Lind_j r}{\Lind_k r}
= \sum_{1\le h,m\le d} \bar{u}_{hj}c_{hm}u_{mk}
= (U^*CU)_{jk}
\]
and vectors $\Lava_j r, \Lava_k r$ are ortogonal.

For all $j$ with $1\le j\le d$, denote  $v_j$ the normalised 
vector $\Lava_j r/\left\Vert \Lava_j r\right\Vert^2$, orthogonal
to $r$. Clearly we have
\begin{eqnarray*}
\lim_{t\to 0^+}\sum_{\ell=1}^k t^{-1}
\lambda_\ell(t)\braket{v_j}{\eava_\ell(t)v_k} 
& = & \sum_{\ell=1}^k \lambda_\ell^\prime(0)
 \braket{v_j}{\eava_\ell(0)v_k} \\
& = & \braket{v_j}{\phiava(D) v_k} \\
& = & \sum_{\ell=1}^d
\braket{v_j}{\ketbra{\Lava_\ell r}{\Lava_\ell r}v_k}
\end{eqnarray*}
for all $j,k$. Therefore $\lambda_\ell^\prime(0)=0$ for
all $\ell=d+1,\dots,k$, $\lambda_\ell^\prime(0) = 
\left\Vert \Lava_\ell r\right\Vert^2$ 
for all $\ell=1,\dots, d$ and $E_\ell(t)$ converges to the 
orthogonal projection onto $v_\ell$ for all $\ell=1,\dots, d$.
{\hfill $\square$}

\begin{lemma}\label{lem:ep-lem-Gterms}
The following limits hold:
\begin{eqnarray*}
\lim_{t\to 0^+} 
\Tr{\ketbra{(\unit\otimes G)r}{r}
\left(\log\deava_t-\log\deind_t\right)} 
& = &  0 \\
\lim_{t\to 0^+} 
\Tr{\ketbra{r}{(\unit\otimes G)r}
\left(\log\deava_t-\log\deind_t\right)} 
& = &  0 \\
\lim_{t\to 0^+} 
\Tr{\ketbra{(G\otimes \unit)r}{r}
\left(\log\deava_t-\log\deind_t\right)} 
& = & 0  \\
\lim_{t\to 0^+} 
\Tr{\ketbra{r}{(G\otimes\unit)r}
\left(\log\deava_t-\log\deind_t\right)} 
& = & 0 
\end{eqnarray*}
\end{lemma}

\noindent{\it Proof.} Clearly
\[
\Tr{\ketbra{(\unit\otimes G)r}{r}
\left(\log\deava_t-\log\deind_t\right)} 
= \left\langle \left(\log\deava_t-\log\deind_t\right)r,
(\unit\otimes G)r\right\rangle.
\]
Writing $\left(\log\deava_t-\log\deind_t\right)r $ as
\[
 \log(\lambda_0(t)) \left(\eava_0(t)r - \eind_0(t)r\right) 
+ \sum_{\ell=1}^k\log(\lambda_\ell(t) )
\left(\eava_\ell(t)r - \eind_\ell (t)r\right)
\]
we start noting that, for $t\to 0^+$, the first term vanishes because 
$\lambda_0(t)$ converges to $1$. The other terms also vanish
because $\eava_\ell(t)r$  and $\eind_\ell (t)r$ converge to $0$ 
for all $\ell\ge 1$ by (\ref{eq:D-eigenvalue1}) and are 
infinitesimal in norm of order $t$ or higher by analyticity. 
Therefore, since $\lambda_\ell(t)$ goes to $0$ polynomially, 
as $t^{m_\ell}$ with $m_\ell\ge 1$, say, we have
\[
\left\Vert \log(\lambda_\ell(t)) \eava_\ell(t)r \right\Vert 
\le c\, t\left| \log(\lambda_\ell(t)) \right|, \qquad
\left\Vert \log(\lambda_\ell(t)) \eind_\ell(t)r \right\Vert 
\le c\, t\left| \log(\lambda_\ell(t)) \right| 
\]
for some constant $c$ and $t$ small enough. 
This proves the first identity. 

The other follow by repeating the above argument. 
{\hfill $\square$}

\noindent{\it Proof.} {\it (of Theorem \ref{th:ep-formula}) } 
The above Lemma \ref{lem:ep-lem-Gterms} and 
(\ref{eq:ep-limit02}) show that it suffices to 
compute the limit for $t\to 0^+$ of
\begin{equation}\label{eq:ep-limit03}
\Tr{\left(\phiava(D)-\phiind(D)\right)
\left(\log\deava_t-\log\deind_t\right)}, 
\end{equation}

Note that, since supports of $\deava_t$ and $\deind_t$ 
are equal, we have 
\[
\sum_{\ell= 0}^k\eava_\ell(t) =\sum_{\ell= 0}^k\eind_\ell(t)
\] 
therefore 
\[
\sum_{\ell=0}^k 
\Tr{\left(\phiava(D)-\phiind(D)\right)\log(t)
\left(\eava_\ell(t)-\eind_\ell(t)\right)}=0.  
\]
Subtracting this from (\ref{eq:ep-limit03}), we can write (\ref{eq:ep-limit03}) as 
\[
\sum_{\ell =0}^k\Tr{\left(\phiava(D)-\phiind(D)\right)
\log\left(\frac{\lambda_\ell(t)}{t}\right)
\left(\eava_\ell(t)-\eind_\ell(t)\right)}.  
\]
Now, the term with $\ell=0$ vanishes for $t$ going to $0$ since 
the logarithm diverges as $\log(t)$ but 
\[
\Tr{\left(\phiava(D)-\phiind(D)\right)
\left(\eava_0(t)-\eind_0(t)\right)}
\]
goes to $0$ (both $\eava_0(t)-\eind_0(t)$ 
converge to $D$, a one-dimensional projection orthogonal to 
the support of $\phiava(D)$ and $\phiind(D)$\,) and the order 
of infinitesimal is at least $t$ by analyticity. 

By Lemma \ref{lem:deava-deind-1st}, $\log(\lambda_\ell(t)/t)$ 
converges to $\log\left\Vert \Lava_\ell\, r\right\Vert^2
=\log\left\Vert \Lava_\ell\, r\right\Vert^2$ and each 
$\eava_\ell(t)$ (resp. $\eind_\ell(t)$) also converges to 
a spectral projection of $\phiava(D)$ (resp. $\phiind(D)$).  
This completes the proof of Theorem \ref{th:ep-formula}.
{\hfill $\square$}

\section{Supports of forward and backward states}
\label{sect:support-problems}

In this section we prove a couple of characterisations of 
the support projection of a pure state evolving under the 
action of a QMS that turn out to be helpful for determining  
the supports of forward and backward densities.

\begin{theorem}\label{th:supp-state}
Let $(\Tt_t)_{t\ge 0}$ be a norm continuous QMS on $\bo{\h}$ 
with generator $\Ll$  as in (\ref{eq:GKSL}) and let $P_t = 
\hbox{\rm e}^{tG}$. For all unit vector 
$u\in\h$ and all $t\ge 0$, the support projection of the state 
$\Tt_{*t}(\ketbra{u}{u})$ is the closed linear span of $P_t u$ 
and vectors 
\begin{equation}\label{eq:supp-state}
P_{s_1}L_{\ell_1}P_{s_{2}-s_{1}}L_{\ell_2}P_{s_3-s_2} \dots 
P_{s_{n}-s_{n-1}}L_{\ell_n}P_{t-s_n}u
\end{equation}
for all $n\ge 1$, $0\le s_1 \le s_{2}\le \dots \le s_{n}\le t$ 
and $\ell_1,\dots,\ell_n\ge 1$.
\end{theorem}

\noindent{\it Proof.} For all $t>0$, differentiating with respect 
to $s$ we have
\begin{eqnarray*}
\frac{d}{ds}\Tt_{*s}\left(P_{t-s}\ketbra{u}{u}P_{t-s}^*\right)
& = & \sum_{\ell\ge 1}\Tt_{*s}
\left(\ketbra{ P_{t-s}L_\ell u}{ P_{t-s}L_\ell u}\right).
\end{eqnarray*}
Integrating on $[0,t]$ we find 
\[
\Tt_{*t}\left(\ketbra{u}{u}\right)
=\ketbra{P_t u}{P_t u} 
+ \sum_{\ell\ge 1}\int_0^t 
\Tt_{*s}\left(\ketbra{L_\ell  P_{t-s}u}{ L_\ell P_{t-s}  u}\right) 
\hbox{\rm d}s.
\]
Iterating yields 
\begin{eqnarray}\label{eq:QMS-iteration}
& & \Tt_{*t}\left(\ketbra{u}{u}\right)
= \ketbra{P_t u}{P_t u} \\
&  & +\sum_{n\ge 1} \sum_{\ell_1,\dots\ell_n\ge 1}
\int_0^t  \hbox{\rm d}s_n \dots \int_0^{s_2} \hbox{\rm d}s_1
\ketbra{u_{t,s_n,\dots,s_1,\ell_1,\dots,\ell_n}}{ u_{t,s_n,\dots,s_1,\ell_1,\dots,\ell_n}}\nonumber
\end{eqnarray}
where $u_{t,s_n,\dots,s_1,\ell_1,\dots,\ell_n}$ is the vector 
given by (\ref{eq:supp-state}). 

Any $v\in\h$, orthogonal to the support of the state 
$ \Tt_{*t}\left(\ketbra{u}{u}\right)$ satisfies 
$\braket{v}{ \Tt_{*t}\left(\ketbra{u}{u}\right)v}=0$. 
Therefore, since all the terms in (\ref{eq:QMS-iteration}) 
are positive operators, it turns out that $v$ must be 
orthogonal to all vectors $P_t u$ and all the iterated 
integrals
\[
\int_0^t  \hbox{\rm d}s_n \dots \int_0^{s_2} \hbox{\rm d}s_1
\left|\braket{v}{P_{s_1}L_{\ell_1}P_{s_{2}-s_{1}}
L_{\ell_2}P_{s_3-s_2} \dots 
P_{s_{n}-s_{n-1}}L_{\ell_n}P_{t-s_n}u}\right|^2 
\] 
vanish. It follows then, from the time continuity of the 
integrands, that $v$ must be orthogonal also to all 
the vectors of the form (\ref{eq:supp-state}) and 
the proof is complete.
{\hfill $\square$}

We now give another characterisation of the support 
of $\Tt_{*t}(\ketbra{u}{u})$ in terms of $P_t$,  
non-commutative polynomials in $L_\ell$ and their 
multiple commutators with $G$. Denote 
$\delta_G^0(L_\ell) = L_\ell, \ 
\delta_G(L_\ell) = \left[ G,L_\ell\right], \
\delta_G^2(L_\ell) = \left[ G,\left[ G,L_\ell\right]\, \right], ...$

\begin{theorem}\label{th:supp-state-differential}
Let $(\Tt_t)_{t\ge 0}$ be a norm continuous QMS on 
${\mathcal{B}}(\h)$ with generator $\Ll$  as in 
(\ref{eq:GKSL}) and let $P_t = \hbox{\rm e}^{tG}$. 
For all unit vector $u\in\h$ and all $t > 0$, the support 
projection of the state $\Tt_{*t}(\ketbra{u}{u})$ is the 
linear manifold $P_t \,\mathcal{S}(u)$ where 
$\mathcal{S}(u)$ is the closure of linear span of $u$ and
\begin{equation}\label{eq:supp-state-diff}
\delta_G^{m_1}(L_{\ell_1})\delta_G^{m_2}(L_{\ell_2}) 
\cdots \delta_G^{m_n}(L_{\ell_{n}})u
\end{equation}
for all $n\ge 1$, $m_1,\dots,m_n\ge 0$ and $\ell_1,
\dots,\ell_n\ge 1$.
\end{theorem}

\noindent{\it Proof.} Let $v$ be a vector orthogonal to the 
suport of $\Tt_{*t}(\ketbra{u}{u})$. Differentiating
\[
\braket{v}{P_{s_1}L_{\ell_1}P_{s_{2}-s_{1}}
L_{\ell_2}P_{s_3-s_2} \dots 
P_{s_{n}-s_{n-1}}L_{\ell_n}P_{t-s_n}u}=0
\]
$m_k$ times with respect to $s_k$ for all $k$, we 
find that $v$ is also orthogonal to $P_t \,\mathcal{S}(u)$.

Conversely, if $v\in\h$ is orthogonal to  $P_t \,\mathcal{S}(u)$, 
then the analytic function 
\[
(s_1,\dots,s_n)\to 
\braket{v}{P_{s_1}L_{\ell_1}P_{s_{2}-s_{1}}
L_{\ell_2}P_{s_3-s_2} \dots 
P_{s_{n}-s_{n-1}}L_{\ell_n}P_{t-s_n}u},
\]
as well as its extension to $\mathbb{C}^n$
\[
(z_1,\dots,z_n)\to 
\braket{v}{P_{z_1}L_{\ell_1}P_{z_{2}-z_{1}}
L_{\ell_2}P_{z_3-z_2} \dots 
P_{z_{n}-z_{n-1}}L_{\ell_n}P_{t-z_n}u},
\]
has all partial derivatives at $z_1=\dots =z_n=t$ equal to $0$. 
Thus it is identically equal to $0$ and $v$ is orthogonal 
to the support of $\Tt_{*t}(\ketbra{u}{u})$.
{\hfill $\square$}

\begin{corollary}\label{cor:G-invariant}
Let $(\Tt_t)_{t\ge 0}$ be a norm continuous QMS on 
${\mathcal{B}}(\h)$ with generator $\Ll$  as in 
(\ref{eq:GKSL}) and let $P_t = \hbox{\rm e}^{tG}$. 
For all unit vector $u\in\h$ the support projection 
of the state $\Tt_{*t}(\ketbra{u}{u})$ is independent 
of $t$, for $t > 0$, if and only if the linear manifold 
$\mathcal{S}(u)$ is $G$-invariant.
\end{corollary}

\noindent{\it Proof.} 
For all $u\in\h$, $S(u)$ is $L_\ell$-invariant for all $\ell\ge 1$ 
because $\delta_G^0(L_\ell)=L_\ell$. If it is also $G$-invariant, 
then it is also $P_t$-invariant for all $t\ge 0$ since 
$P_t=\sum_{n\ge 0} t^nG^n/n!$ and supports of states 
$\Tt_{*t}(\ketbra{u}{u})$ coincide with $S(u)$  for all $t>0$ by 
Theorem \ref{th:supp-state-differential}. 

Conversely, if  the support projection of 
$\Tt_{*t}(\ketbra{u}{u})$ is independent of $t$, then 
$P_{t}\mathcal{S}(u)=\mathcal{S}(u)$ for all $t\ge 0$, 
by continuity of $P_t$ at $t=0$. Differentiating at $t=0$  
we find then $G\mathcal{S}(u)\subseteq \mathcal{S}(u)$.
{\hfill $\square$}

\begin{theorem}\label{th:support-state2}
Let $\Tt$ be a QMS with generator $\Ll$ as in Theorem  
\ref{th-special-GKSL} and suppose that $\rho^{1/2}\theta G^*\theta=G\rho^{1/2}$. The following conditions are equivalent:
\begin{itemize}
\item[(a)]  the closed linear spans of  
$\set{L_\ell \rho^{1/2}\,\mid\, \ell\ge 1}$ and 
$\set{ \rho^{1/2}\theta L_\ell^*\theta \,\mid\, \ell\ge 1}$ 
in the Hilbert space of Hilbert-Schmidt operators on $\h$ 
coincide,
\item[(b)] the forward and backward states $\deava_t$ 
and $\deind_t$ have the same support.
\end{itemize}
\end{theorem}

\noindent{\it Proof.} Putting 
$\overrightarrow{\Tt}_t = I\otimes \Tt_t$ and 
$\overleftarrow{\Tt}_t = \Tt_t  \otimes I$, we 
define the forward and backward QMS 
$\overrightarrow{\Tt}$ and $\overleftarrow{\Tt}$ 
on $\bo{\h}\otimes\bo{\h}$. Their generators can 
be written in a special GKSL representation, with respect 
to the faithful normal invariant state $\rho\otimes \rho$ 
by means of operators $\overrightarrow{G}=\unit\otimes G$, 
$\overrightarrow{L}_\ell=\unit\otimes L_\ell$ and 
$\overleftarrow{G}= G\otimes \unit$, 
$\overleftarrow{L}_\ell= L_\ell\otimes\unit$. 
Denote $(\overrightarrow{P}_t)_{t\ge 0}$ and 
$(\overleftarrow{P}_t)_{t\ge 0}$ the semigroups 
on $\h\otimes \h$ generated by $\overrightarrow{G}$ 
and $\overleftarrow{G}$ respectively.

By Theorem \ref{th:supp-state}, it suffices to show that 
condition (a) holds if and only if the closed linear spans 
in $\h\otimes\h$ of the sets 
\begin{eqnarray} 
& & \kern-36truept\overrightarrow{P}_{t}r, \quad
\overrightarrow{P}_{s_1}\overrightarrow{L}_{\ell_1}
\overrightarrow{P}_{s_{2}-s_{1}}\overrightarrow{L}_{\ell_2}
\overrightarrow{P}_{s_3-s_2} \dots 
\overrightarrow{P}_{s_{n}-s_{n-1}}\overrightarrow{L}_{\ell_n}
\overrightarrow{P}_{t-s_n}r \label{eq:supp-state-fw}\\
& & \kern-36truept\overleftarrow{P}_{t}r, \quad
\overleftarrow{P}_{s_1}\overleftarrow{L}_{\ell_1}
\overleftarrow{P}_{s_{2}-s_{1}}\overleftarrow{L}_{\ell_2}
\overleftarrow{P}_{s_3-s_2} \dots 
\overleftarrow{P}_{s_{n}-s_{n-1}}\overleftarrow{L}_{\ell_n}
\overleftarrow{P}_{t-s_n}r \label{eq:supp-state-bw}
\end{eqnarray}
for all $n\ge 1$, $0\le s_1 \le s_{2}\le \dots \le s_{n}\le t$ 
and $\ell_1,\dots,\ell_n\ge 1$ coincide.

Let $w=\sum_{\alpha,\beta} w_{\beta \alpha} 
e_\alpha\otimes e_\beta$ be a vector in $\h\otimes \h$.
Note that $\left\Vert{w}\right\Vert^2=\sum_{\alpha,\beta} 
\left|w_{\beta\alpha}\right|^2$, therefore the matrix 
$(w_{\beta\alpha})_{\alpha,\beta\ge 1}$ defines 
a Hilbert-Schmidt operator $W$ on $\h$ with 
$w_{\beta\alpha}=\braket{e_\alpha}{W e_\beta}$.
The vector $w$ is orthogonal to $(X\otimes\unit )r$
 if and only if 
\[
0 = \sum_{j,\alpha,\beta}\rho_j^{1/2}
\braket{(X\otimes\unit )e_j\otimes e_j}
{e_\alpha\otimes e_\beta}w_{\beta\alpha} 
= \sum_{j,\alpha}\rho_j^{1/2}
\braket{Xe_j}{e_\alpha}\braket{e_j}{W e_{\alpha}}
\]
i.e.
\begin{eqnarray*}
0 &=&\sum_{j,\alpha}\rho_j^{1/2}
\braket{e_\alpha}{\theta X\theta e_j}\braket{e_j}{W e_{\alpha}} \\
& = & \sum_{j,\alpha}
\braket{\rho^{1/2}\theta X^*\theta e_\alpha}{e_j}
\braket{e_j}{W e_{\alpha}} \\
& = & \tr{\left(\rho^{1/2}\theta X^*\theta\right)^*W}
\end{eqnarray*}
namely $\rho^{1/2}\theta X^*\theta$ is orthogonal to 
$W$ in Hilbert-Schmidt operators on $\h$. In a similar way, 
a straightforward computation shows that $w$ is orthogonal 
to $(\unit \otimes X)r$ if and only if $X\rho^{1/2}$ is orthogonal 
to $W$  in Hilbert-Schmidt operators on $\h$.

Since $\rho^{1/2}\theta G^*\theta = G\rho^{1/2}$, 
by induction we have immediately 
 $\rho^{1/2}\theta G^{*k}\theta = G^k\rho^{1/2}$
for all $k\ge 0$ and then 
\[
P_t\rho^{1/2}=\sum_{k\ge 0}\frac{t^k}{k!}G^k\rho^{1/2}
=\sum_{k\ge 0}\frac{t^k}{k!}\rho^{1/2}\theta G^{*k}\theta
=\rho^{1/2}\theta P_t^*\theta.
\]
Thus $w$ is orthogonal to $\overrightarrow{P}_{t}r$ if and only if 
the Hilbert-Schmidt operator $W$ is orthogonal to 
$P_t\rho^{1/2}= \rho^{1/2}\theta P_t^*\theta$ 
namely $w$ is orthogonal to  $\overleftarrow{P}_{t}r$.
Moreover, $w$ is orthogonal to the second vector in 
(\ref{eq:supp-state-fw}) given by 
\[
\left(\unit\otimes(P_{s_1} L_{\ell_1}
P_{s_{2}-s_{1}} L_{\ell_2}
P_{s_3-s_2} \dots P_{s_{n}-s_{n-1}}L_{\ell_n}
P_{t-s_n})\right)r
\]
if and only if $W$ is orthogonal to 
\[
P_{s_1} L_{\ell_1}
P_{s_{2}-s_{1}} L_{\ell_2}
P_{s_3-s_2} \dots P_{s_{n}-s_{n-1}}L_{\ell_n}
P_{t-s_n}\rho^{1/2}
\]
namely $W$ is orthogonal to 
\[
\rho^{1/2}\theta(P_{s_1} L_{\ell_1}
P_{s_{2}-s_{1}} L_{\ell_2}
P_{s_3-s_2} \dots P_{s_{n}-s_{n-1}}L_{\ell_n}
P_{t-s_n})^*\theta  
\]
namely $w$ is orthogonal to the second vector in 
(\ref{eq:supp-state-bw}).
{\hfill $\square$}


\begin{proposition}\label{prop:supp-HS=derivatives}
The following conditions are equivalent:
\begin{itemize}
\item[(a)] the closures of the linear spans of  
$\set{L_\ell \rho^{1/2}\,\mid\, \ell\ge 1}$ and 
$\set{ \rho^{1/2}\theta L_\ell^*\theta \,\mid\, \ell\ge 1}$ 
in the Hilbert space of Hilbert-Schmidt operators on $\h$ 
coincide,
\item[(b)] the supports of  $\phiava(D)$ and $\phiind(D)$ coincide.
\end{itemize}
\end{proposition}

\noindent{\it Proof.} Let $w = \sum_{\alpha,\beta} w_{\beta\alpha}
\,\theta e_\alpha\otimes e_\beta$ be a vector in $\h\otimes\h$ orthogonal to $r$ and let $W$ be the Hilbert-Schmidt operator 
$\h\otimes\h$ with $w_{\beta\alpha}=
\braket{e_\alpha}{W e_\beta}$.
Straightforward computations yield
\begin{eqnarray*}
\phiava(D)w & = &   \sum_{\ell,\alpha,\beta}
w_{\beta \alpha} 
\braket{(\unit\otimes L_\ell) r}{\theta e_\alpha\otimes e_\beta} 
(\unit\otimes L_\ell) r, \\
\phiind(D) w
& = & \sum_{\ell,\alpha,\beta}
w_{\beta\alpha}
\braket{(L_\ell\otimes \unit) r}{\theta e_\alpha\otimes e_\beta} 
 (L_\ell\otimes \unit)r.
\end{eqnarray*}
If $\phiava(D)w=0$, since the vector $r$ is separating for $\unit\otimes \bo{\h}$, we have  
\[
\sum_{\ell,\alpha,\beta}
w_{\beta \alpha} 
\braket{(\unit\otimes L_\ell) r}{\theta e_\alpha\otimes e_\beta} 
(\unit\otimes L_\ell) 
=   \sum_{\ell,\alpha,\beta}
w_{\beta \alpha} \rho_{\alpha}^{1/2}
\braket{ L_\ell e_\alpha}{e_\beta}(\unit\otimes L_\ell)  = 0.
\]
namely, by the linear independence of the $L_\ell$, 
\[
0=\sum_{\alpha,\beta}
w_{\beta \alpha} \rho_{\alpha}^{1/2}
\braket{ L_\ell e_\alpha}{e_\beta} 
= \sum_{\alpha,\beta}
w_{\beta \alpha} 
\braket{ L_\ell \rho^{1/2}e_\alpha}{e_\beta} 
= \sum_{\alpha}
\braket{ L_\ell \rho^{1/2}e_\alpha}{We_\alpha} 
\]
for all $\ell \ge 1$. Therefore $\phiava(D)w=0$ if and only if 
$ \tr{(L_\ell\rho^{1/2})^*W}=0$.

We can show that 
$\phiind(D)w=0$ if and only if 
$ \tr{(\rho^{1/2}\theta L_\ell^*\theta)^*W}=0$
in the same way.
It follows that   
$\set{L_\ell \rho^{1/2}\,\mid\, \ell\ge 1}$ and 
$\set{ \rho^{1/2}\theta L_\ell^*\theta \,\mid\, \ell\ge 1}$ 
in the Hilbert space of Hilbert-Schmidt operators on $\h$ 
have the same orthogonal and the equivalence of (a) 
and (b) is clear. 
{\hfill $\square$}

\section{Examples}\label{Sect:examples}

In this section we collect three examples illustrating our entropy production formula.  The antiunitary $\theta$ will always be 
conjugation with respect to the chosen basis of $\h$.

\subsection{Trivial cycle on an $n$-level system}

Consider the QMS on $\bo{\mathbb{C}^n}$ $(n\ge 3$) 
generated by
\[
\Ll(x) = \lambda\, S^* x S + \mu\, SxS^* - x 
+ \mi  [H,x]
\]
where $S$ is the unitary right shift defined on the orthonormal 
basis $(e_j)_{0\le j\le n-1}$ of $\mathbb{C}^n$ by 
$S e_ {j} = e_{j+1}$ (the sum must be understood mod $n$), 
$\lambda,\mu>0$. The Hamiltonian $H$ is a real matrix 
which is diagonal in this basis.

This QMS may arise in the stochastic (weak coupling) limit of a 
three-level system dipole-type interacting with two reservoirs 
at different temperatures under the generalised rotating wave approximation. The parameters $\lambda,\mu$ are related 
to the temperatures of the reservoirs and $\lambda=\mu$ if 
the temperatures coincide.
Its structure  is clear:
\begin{enumerate}
\item $\rho= \unit/n\,$ is a faithful invariant state, therefore 
the QMS  commutes with the trivial modular group,
\item $d=2$, and $L_1=\lambda^{1/2}S$, $L_2
=\mu^{1/2} S^*$, together with  $G=-2^{-1}\unit -\mi H$
give a special GKSL representation of $\Ll$,
\item we have $\rho^{1/2}\theta G^*\theta = \rho^{1/2}G 
=G\rho^{1/2}$, 
\item quantum detailed balance conditions are satisfied if 
and only if $\lambda=\mu$ since 
\[
\left[\begin{array}{c}
\rho^{1/2}\theta L_1^*\theta \\
\rho^{1/2}\theta L_1^*\theta
\end{array}\right]
=\left[\begin{array}{cc}
0 & (\mu/\lambda)^{1/2} \\
(\lambda/\mu)^{1/2} & 0 
\end{array}\right]
\left[\begin{array}{c}
L_1\rho^{1/2}  \\
L_2 \rho^{1/2}
\end{array}\right]
\]
\end{enumerate}
and the above matrix is unitary if and only if $\lambda=\mu$.

A complete study of the qualitative behaviour of this QMS can be 
done by applying our methods in \cite{FFRR-LNM1882}. 

The assumption ({\bf FBS}) is immediately checked applying 
Theorem \ref{th:support-state2} (a) because the linear  
spans of both set of operators coincide with the Abelian 
algebra generated by the shift $S$, namely the algebra of 
$n\times n$ circulant matrices.

The entropy production is easily computed applying 
our formula (\ref{eq:ep-formula}). Indeed 
\begin{eqnarray*}
\phiava(D) & = & \frac{\lambda}{n}\sum_{j,k=0}^{n-1} 
\ketbra{e_j\otimes e_{j+1}}{e_k\otimes e_{k+1}}
+   \frac{\mu}{n}\sum_{j,k=0}^{n-1} 
\ketbra{e_j\otimes e_{j-1}}{e_k\otimes e_{k-1}}  \\
\phiind(D) & = & \frac{\lambda}{n}\sum_{j,k=0}^{n-1} 
\ketbra{e_{j+1}\otimes e_j}{e_{k+1}\otimes e_k}
+   \frac{\mu}{n}\sum_{j,k=0}^{n-1} 
\ketbra{e_{j-1}\otimes e_j}{e_{k-1}\otimes e_k}  
\end{eqnarray*}
where  sums $j\pm 1, k\pm 1$ are modulo $n$. A quick inspection 
shows that, denoting $\psi_{+}, \psi_{-}$ the unit vectors 
\[
\psi_{+} = \frac{1}{\sqrt{n}}\sum_{j=0}^{n-1}
e_j\otimes e_{j+1}, \qquad
\psi_{-}=\frac{1}{\sqrt{n}}\sum_{j=0}^{n-1}
e_j\otimes e_{j-1},
\]
we have $\left\langle \psi_{-},\psi_{+}\right\rangle=0$ and
\[
\phiava(D) = \lambda \ketbra{\psi_{+}}{\psi_{+}} 
+ \mu \ketbra{\psi_{-}}{\psi_{-}}, \qquad
\phiind(D) =\lambda \ketbra{\psi_{-}}{\psi_{-}} 
+ \mu\ketbra{\psi_{+}}{\psi_{+}}. 
\]
It follows that
\begin{eqnarray*}
\phiava(D) - \phiind(D) & = & (\lambda -\mu)
 \left( \ketbra{\psi_{+}}{\psi_{+}}-\ketbra{\psi_{-}}{\psi_{-}}\right) \\
\log\left(\phiava(D)\right) - \log\left(\phiind(D)\right)
 & = & \log\left(\frac{\lambda}{\mu} \right)
 \left( \ketbra{\psi_{+}}{\psi_{+}}-\ketbra{\psi_{-}}{\psi_{-}}\right) 
\end{eqnarray*}
and the entropy production is
\[
\frac{\lambda-\mu}{2}\log\left(\frac{\lambda}{\mu} \right).
\]
Therefore, the entropy production is non zero if and only if $\lambda\not=\mu$ since there is  a ``current'' determined by 
different intensities in ``raising'' ($e_{j}\to e_{j+1}$) and 
``lowering'' ($e_{k}\to e_{k-1}$) transitions. 

Note that this entropy production coincides with the entropy 
production of the classical QMS obtained by restriction to the 
commutative subalgebra of diagonal matrices.

\subsection{Generic QMS} 

Generic QMS arise in the stochastic limit of a open discrete 
quantum system with generic Hamiltonian, interacting with 
Gaussian fields through a dipole type  interaction (see 
\cite{AcLuVo,CaFaHa}).  Here, for 
simplicity, the system space is finite-dimensional 
$\h=\mathbb{C}^n$ with orthonormal basis 
$(e_j)_{0\le j\le n-1}$, the operators $L_\ell$, in this 
case labeled by a double index $(\ell,m)$ with 
$\ell\not=m$, are 
\[
L_{\ell m} = \gamma_{\ell m}^{1/2}\ketbra{e_m}{e_\ell}
\]
where are $\gamma_{\ell m}\ge 0$  positive constants and 
the effective Hamiltonian $H$ is a self-adjoint operator 
diagonal in the given basis whose explicit form is not 
needed here because it does not affect the entropy 
production. The generator $\Ll$ is 
\begin{eqnarray}
\label{Equation-GeneratorGenericMarkovSemigrp} 
\Ll(x) =  \mi [H,x]+\frac{1}{2} \sum_{\ell\not=m} 
\left(-L_{\ell m}^*L_{\ell m} x + 2 L_{\ell m}^* x L_{\ell m}
- x L_{\ell m}^*L_{\ell m}\right), 
\end{eqnarray}
therefore 
\[
G=-\frac{1}{2}\sum_{\ell\not= m} L_{\ell m}^*L_{\ell m} - \mi H
= -\frac{1}{2}\sum_\ell \left(\sum_{\{m\,\mid\,m\not=\ell\,\}} 
\gamma_{\ell m}\right) \ketbra{e_\ell}{e_\ell} -\mi H
\]
is diagonal in the given basis and the condition $\rho^{1/2}\theta 
G^*\theta = G\rho^{1/2}$ holds. Moreover, for any given 
faithful normal state (even if it is \emph{not} an invariant state) 
$\rho=\sum_{j=0}^n\ketbra{e_j}{e_j}$ we have
\[
L_{\ell m}\rho^{1/2} 
= \rho_\ell^{1/2}\gamma_{\ell m}^{1/2}\ketbra{e_m}{e_\ell}, 
\qquad
\rho^{1/2}\theta L_{\ell m}^*\theta =
\rho_\ell^{1/2}\gamma_{\ell m}^{1/2}\ketbra{e_\ell}{e_m}.
\]
It follows that the linear span of operators  
$L_{\ell m}\rho^{1/2}$ coincides with the linear span of 
operators $\rho^{1/2}\theta L_{\ell m}^*\theta$ 
if and only if $\gamma_{\ell m}>0$ implies 
$\gamma_{m\ell}>0$ for all $\ell,m$. 
Under this assumption ({\bf FBS}) clearly holds.

The restriction of $\Ll$ to the algebra of diagonal matrices 
coincides with the generator of a time continuous Markov 
chain with states ${0,1,\dots, n-1}$ and jump rates 
$\gamma_{\ell m}$. 
As a consequence, if $\gamma_{\ell m}>0$ implies 
$\gamma_{m\ell}>0$ for all $\ell,m$ the classical 
time-continuous Markov chain can be realised as a 
union of its irreducible classes each one of them admitting 
a unique strictly positive invariant probability density. Any 
convex combination of these probability densities with 
all non-zero coefficients yields and invariant probability 
density $(\rho_j)_{0\le j\le n-1}$ for the whole Markov 
chain with $\rho_j>0$ for all $j$. 
It is easy to check that the diagonal matrix with eigenvalues 
$(\rho_j)_{0\le j\le n-1}$ is an invariant state for the 
quantum Markov semigroup generated by $\Ll$.

Straightforward computations give the following formulae: 
\begin{eqnarray*}
\phiava
(D) & = &\sum_{\{\,(\ell,m)\,\mid\, \gamma_{\ell m}>0\,\}}
\rho_\ell \,\gamma_{\ell m} \,
\ketbra{e_\ell\otimes e_m}{e_\ell\otimes e_m} \\
\phiind(D) & = &\sum_{\{\,(\ell,m)\,\mid\, \gamma_{\ell m}>0\,\}}
\rho_m \gamma_{m\ell}\,\ketbra{e_\ell\otimes e_m}{e_\ell\otimes e_m}
\end{eqnarray*}
Therefore the entropy production is 
\[
\frac{1}{2}\sum_{\{\,(\ell,m)\,\mid\, \gamma_{\ell m}>0\,\}}
\left({\rho_\ell\, \gamma_{\ell m}}-{\rho_m\gamma_{m\ell}} \right) 
\log\left( \frac{\rho_\ell\, \gamma_{\ell m}}{\rho_m\gamma_{m\ell}}\right).
\]
This formula shows immediately that the entropy 
production is zero
if and only if the classical detailed balance condition 
${\rho_\ell\, \gamma_{\ell m}}={\rho_m\gamma_{m\ell}}$
for all $\ell, m$ holds. Here again, entropy production coincides 
with the entropy production of the classical QMS obtained 
by restriction to the commutative subalgebra of diagonal 
matrices. Moreover, it is not difficult to show that, 
if there is a $\gamma_{\ell m}>0$ with $\gamma_{m\ell}=0$ 
and the classical Markov chain is irreducible, the invariant 
state is faithful but the entropy production is infinite.

\subsection{Two-level system}\label{ex:two-level}

Let $\Tt$ be the QMS on $\bo{\mathbb{C}^2}$ with generator 
$\Ll$ represented in a GKSL form with 
\[
L_1=\ketbra{e_1}{e_2}, \quad L_2=\ketbra{e_2}{e_1}, 
\quad H =  \mi \kappa\left( \ketbra{e_2}{e_1}-\ketbra{e_1}{e_2}\right), \quad \kappa\in\mathbb{R}-\{0\}.
\]
The normalised trace $\rho=\unit/2$ is a faithful invariant 
state and the above operator give a special GKSL 
representation of $\Ll$.

The semigroup $\Tt$ satisfies the SQDB condition by 
Theorem \ref{th:SQDB}. Indeed 
\[
\rho^{1/2}L_1^* = L_2\, \rho^{1/2}, \qquad 
\rho^{1/2}L_2^* = L_1\, \rho^{1/2}
\] 
so that we can choose as self-adjoint unitary  in 
(\ref{sqdb-cond}) the flip $u e_1 =e _2,\, ue_2= e_1$. 

The SQBD-$\Theta$ condition, however, does not hold 
because 
\[
\rho^{1/2}\theta G^* \theta  - G \rho^{1/2} 
=  2\mi  H \rho^{1/2} \not=0.
\]

Computing $\left[ G, L_1\right] 
= \left[ G, L_2\right]  
=   \kappa\left(\ketbra{e_1}{e_1} - \ketbra{e_2}{e_2}\right)$ 
and noting that  
\begin{eqnarray*}
(\unit\otimes L_1) r = {e_2\otimes e_1}/{\sqrt{2}}, &\quad&
(\unit\otimes L_2) r = {e_1\otimes e_2}/{\sqrt{2}}, \\
(\unit\otimes \left[ G, L_1\right]) r 
& = & {\kappa(e_1\otimes e_1 - e_2\otimes e_2 )}/{\sqrt{2}}, 
\end{eqnarray*}
by the invertibility of $\unit\otimes P_t$, we find immediately 
that the support of $\deava_t$ is the whole $\mathbb{C}^2\otimes \mathbb{C}^2$ by 
Theorem \ref{th:supp-state-differential}. The support of 
$\deind_t$ is the same since $\deind_t=F\,\deava_t F$ where 
$F$ is the unitary flip $F e_j\otimes e_k = e_k\otimes e_j$. 
Therefore the assumption ({\bf FBS}) holds. 
A simple computation yields
\[
\phiava(D) = \phiind(D)=
\frac{1}{2}\left( \ketbra{e_1\otimes e_2}{e_1\otimes e_2}
+\ketbra{e_2\otimes e_1}{e_2\otimes e_1}\right),
\]
thus the entropy production is zero.

\section{Conclusions and outlook}

We showed that strictly positivity of entropy production characterises 
non equilibrium invariant states of quantum Markov semigroups, 
irrespectively of the chosen notion of quantum detailed balance 
and  commutation with the modular group. Entropy production 
only depends on the completely positive part of the generator 
of a QMS that can be regarded as its truly irreversible part. 

States with finite entropy production form a promising class of 
non equilibrium invariant states. Indeed, they satisfy an operator 
version (Theorem \ref{th:support-state2}) of  the necessary  
condition for finiteness of classical entropy production 
$\gamma_{jk}>0$ if and only if $\gamma_{kj}>0$ where 
$\gamma_{jk}$ are transition rates. Moreover dependence 
of entropy production on the completely positive part of the 
generator of a QMS only might allow us to extend cycle 
decompositions of QMS like those obtained in 
\cite{AcFaQu,BolQue,FFVU12} to QMS non commuting 
with the modular group.
These directions will be explored in forthcoming papers.

\section*{Appendix}

\begin{proposition}\label{prop:onb-theta-invariant}
If the state $\rho$ and $\theta$ commute there exists an 
orthonormal basis $(e_j)_{j\ge 1}$ of $\h$ of eigenvectors 
of $\rho$ that are all invariant under $\theta$.
\end{proposition} 

\noindent{\it Proof.}
Let $(e_j)_{j\ge 1}$ of $\h$ of eigenvectors of $\rho$ and let 
$\rho= \sum_{j\ge 1} \ketbra{e_j}{e_j}$
be a spectral decomposition of $\rho$ with $\rho_j>0$ for 
all $j\ge 1$ because $\rho$ is faithful. Since $\theta$ commutes 
with $\rho$  we have 
$\rho \theta e_j = \theta \rho e_j = \rho_j \theta e_j$,
and eigenspaces of $\rho$ are $\theta$-invariant. 
Now, for each $j$ such that $\theta e_j \not=-e_j$, 
the normalised vector $f_j=(e_j+\theta e_j)/\left\Vert e_j+\theta e_j\right\Vert$  is $\theta$-invariant and is still an 
eigenvector of $\rho$ as well as $f_j=\mi e_j$ if 
$\theta e_j =-e_j$. Noting that scalar products $\braket{f_j}{f_k}$ 
are real, since 
$\braket{f_j}{f_k} = \braket{\theta f_k}{\theta f_j}
=\braket{f_k}{f_j}$,
by a standard Gram-Schmidt orthogonalisation process 
we can find an orthonormal basis of the eigenspace of 
$\rho_j$ of $\theta$-invariant vectors.
{\hfill $\square$}

\section*{Acknowledgements} 
Thanks to Alessandro Toigo for useful discussions and 
a careful reading of the paper.
Financial support from FONDECYT 1120063 and ``Stochastic Analisis Networt'' CONICYT-PIA grant ACT 1112 is gratefully acknowledeged.


\begin{thebibliography}{}
%
%


\bibitem{AcFaQu}
Accardi,~L., Fagnola,~F.,  Quezada,~R.:
Weighted Detailed Balance and Local KMS Condition
for Non-Equilibrium Stationary States, 
{Bussei Kenkyu}  \textbf{97}, (2011) 318-356.


\bibitem{AcLuVo} 
L. Accardi, Y. G. Lu and I. Volovich, 
\textit{Quantum theory and its stochastic 
limit}, Springer-Verlag, Berlin, (2002).

\bibitem{Agarwal}
 Agarwal,~G.S.: Open quantum Markovian systems and the microreversibility. {Z. Physik}  \textbf{258}, (1973) 409--422.

\bibitem{Alicki76}
 Alicki,~R.:  On the detailed balance condition for non-Hamiltonian systems, {Rep. Math. Phys.}, \textbf{10} (1976) 249--258.

\bibitem{Alicki-Lendi}
Alicki,~R., Lendi,~K.:\textit{Quantum Dynamical Semigroups 
and Applications}, \textit{Lecture Notes in Physics} \textbf{286}, 
Springer-Verlag, Berlin 1987.

\bibitem{BolQue}
Bola\~nos, J., Quezada, R.: A cycle decomposition and entropy 
production for circulant quantum Markov semigroups.
{\tt arXiv:1210.6401v1}

\bibitem{HeinzPeter:2003p34}
H.P. Breuer.
\newblock Quantum jumps and entropy production.
\newblock {\em Phys. Rev. A}, {\bf 68} (2003), 032105.

\bibitem{MR2046709}
Callens, I., De~Roeck, W., Jacobs, T.,  Maes, C., Neto{\v{c}}n{\'y}, K.:
\newblock Quantum entropy production as a measure 
of irreversibility.
\newblock {\em Phys. D}, \textbf{187} (2004) 383--391.

\bibitem{CaFaHa}
Carbone, R., Fagnola, F., Hachicha, S.: 
Generic quantum Markov semigroups: 
the Gaussian gauge invariant case. 
{\it Open Syst. Inf. Dyn.} {\bf 14} (2007), 425-444.

\bibitem{Cipriani}
Cipriani, F.: Dirichlet forms and markovian semigroups on 
standard forms of von Neumann algebras.
\textit{J. Funct. Anal.} \textbf{147},  (1997) 259--300.


\bibitem{Qian2003}
Da-Quan Jiang, Min Qian, and Fu-Xi Zhang.
\newblock Entropy production fluctuations of finite 
{M}arkov chains.
\newblock {\em J. Math. Phys.} {\bf 44}, (2003) 4176--4188.

\bibitem{DeFr}
Derezynski, J., Fruboes, R.: Fermi golden rule and open quantum 
systems, in: {\em Open Quantum Systems III - 
Recent Developments}, 
Lecture Notes in Mathematics {\bf 1882}, 
Springer Berlin, Heidelberg (2006), pp. 67–116.

\bibitem{FFRR-LNM1882}
Fagnola, F., Rebolledo, R.: Notes on the Qualitative Behaviour 
of Quantum Markov Semigroups, in:  \textit{Open Quantum 
Systems III - Recent Developments}. Lecture Notes in 
Mathematics \textbf{1882}, Springer Berlin, Heidelberg (2006), 
pp. 161--206.

\bibitem{Fagnola-Rebolledo-QP29}
Fagnola, F., Rebolledo, R.: {From classical to quantum 
entropy production}, in  \textit{Quantum Probability and 
Infinite Dimensional Analysis},  QP-PQ: Quantum Probability 
and White Noise Analysis \textbf{25},  
 World Scientific, Singapore (2010), pp. 245--261.

\bibitem{FFVU07}
Fagnola, F., Umanit\`a, V.: Generators of detailed balance 
quantum Markov semigroups. {\it Inf. Dim. Anal. Quant. Probab. 
Relat. Top.} {\bf 10},  (2007) 335--363.

\bibitem{FFVU}
Fagnola, F., Umanit\`a, V. :
Generators of KMS Symmetric Markov Semigroups on $\bo{\h}$
Symmetry and Quantum Detailed Balance. 
{\it Commun. Math. Phys.} {\bf 298}, (2010) 523--547. 

\bibitem{FFVU12}
Fagnola, F., Umanit\`a, V. :
Generic Quantum Markov Semigroups, Cycle Decomposition 
and Deviation From Equilibrium, {\sl Infin. Dimens. Anal. 
Quantum Probab. Relat. Top.}, {\bf 15}
No. 3 (2012) 1250016 (17 pages).

\bibitem{GoLi}
Goldstein, S., Lindsay, J.M.:  Beurling-Deny condition 
for KMS-symmetric dynamical semigroups,  
{\sl C. R. Acad. Sci. Paris} {\bf 317}, (1993) 1053--1057. 

\bibitem{JaksicPill}
Jak{\v s}i{\'c}, V., Pillet, C.-A.:
\newblock On entropy production in quantum statistical mechanics.
\newblock \emph {Commun. Math. Phys.}
\textbf{217} (2001), (2001) 285--293.  

\bibitem{Kato66}
Kato, T.: {\sl Perturbation theory for linear operators}. 
Springer-Verlag, 1966.

\bibitem{KFGV}
Kossakowski, A., Frigerio, A., Gorini V., Verri, M.:
Quantum detailed balance and KMS condition.
{\it  Comm. Math. Phys.}  {\bf 57},  (1977)  97--110.

\bibitem{MRM}
Maes, C., Redig, F., Van~Moffaert, A.:
\newblock On the definition of entropy production, via examples.
\newblock {\em J. Math. Phys.}, {\bf 41} (2000), 1528--1554. 

\bibitem{Maje}
Majewski,~W.A.: 
The detailed balance condition in quantum statistical mechanics,
{J. Math. Phys.} \textbf{25}, (1984) 614--616. 

\bibitem{MaSt}
Majewski,~W.A., Streater,~R.F.: Detailed balance and quantum dynamical maps,  {\it J. Phys. A: Math. Gen.} \textbf{31},  (1998) 
7981--7995.

\bibitem{Onsager:1931p1766}
Onsager, L.: 
\newblock Reciprocal relations in irreversible processes. I.
\newblock {\em Phys Rev} {\bf 37}, (1931) 405--426.

\bibitem{Partha}  
Parthasarathy,~K.R.: 
\textit{An introduction to quantum stochastic calculus}, 
{Monographs in Mathematics} \textbf{85}, 
Birkh\"auser-Verlag, Basel 1992.

\bibitem{Petz}
Petz,~D.: Conditional expectation in quantum probability, 
in \textit{Quantum Probability and Applications III}.  
Lecture Notes in Mathematics  \textbf{1303} 
Springer, Berlin-Heidelberg-New York 1988, pp. 251–260.

\bibitem{RSv4}
Reed,~M., Simon,~B.:
\textit{Analysis of Operators}, Volume \textbf{IV} of 
\textit{Methods of Modern Mathematical Physics}.
Academic Press, San Diego 1978.

\bibitem{Talkner}
Talkner,~P.: The failure of the quantum regression hypothesis, 
{ Ann. Physics}  \textbf{167} (1986),  390--436.


\end{thebibliography}
\end{document}